\documentclass[12pt]{iopart}
\usepackage{epsfig}
\usepackage{amssymb}

\usepackage{epsfig}
\usepackage{graphicx}
\usepackage{bm}
\usepackage{color}

\newcommand{\pphi}{{\cal P}_{\phi}^{1/2}}
\newcommand{\vppp}{V_0^{\textrm{\tiny(3)}}}
\newcommand{\vpppp}{V_0^{\textrm{\tiny(4)}}}
\def\thebiblio#1{
\begin{center}\bf \large References
\end{center}
\list
{[\arabic{enumi}]}{\settowidth\labelwidth{#1.}\leftmargin\labelwidth
 \advance\leftmargin\labelsep
 \usecounter{enumi}}
 \def\newblock{\hskip .11em plus .33em minus -.07em}
 \sloppy
 \sfcode`\.=1000\relax}

\begin{document}
\title{Trapped Quintessential Inflation in the context of Flux Compactifications}

\author{J.C. Bueno S\'anchez and K. Dimopoulos}

\begin{abstract}
We present a model for quintessential inflation using a string
modulus for the inflaton - quintessence field. The scalar potential
of our model is based on generic non-perturbative potentials arising
in flux compactifications. We assume an enhanced symmetry point (ESP),
which fixes the initial conditions for slow-roll inflation.
When crossing the ESP the modulus becomes temporarily trapped, which
leads to a brief stage of trapped inflation. This is followed by enough
slow roll inflation to solve the flatness and horizon problems.
After inflation, the field rolls down the potential and eventually
freezes to a certain value because of cosmological friction. The latter
is due to the thermal bath of the hot big bang, which is produced
by the decay of a curvaton field. The modulus remains frozen until the
present, when it becomes quintessence.

\end{abstract}

\maketitle
\section{\label{sec:intro}Introduction}
An array of recent observations has ascertained that we live in a
Universe which is engaging into accelerated expansion at present
\cite{all}. The simplest explanation for this observation is to
assume a non-zero cosmological constant, corresponding to vacuum
energy density comparable to the density of matter today. However,
such a value for the cosmological constant is extremely fine-tuned
compared to theoretical expectations \cite{L}. As a result, a
number of alternatives has been suggested in the literature. Many
of the alternative solutions postulate the existence of an unknown
exotic substance, called Dark Energy, whose properties (e.g.
equation of state) are such that it would drive the Universe to
accelerated expansion if it dominates the Universe content (for a
recent review see \cite{ed}). A particular type of such substance,
which fulfils the requirements of dark energy is a potentially
dominated scalar field. Under this hypothesis, the cosmological
constant may be set to zero, as once assumed, so that the extreme
fine-tuning of its value can be avoided. This means that the
scalar field either lies at a metastable minimum of the scalar
potential or is rolling down a slope in the scalar potential
leading to the minimum, which corresponds to zero density. This
latter case has been envisaged some time ago in models of
`dynamical cosmological constant' \cite{early}. With respect to
accounting for dark energy, the scalar field has been named
`quintessence' because it is the fifth element after baryons,
photons, CDM and neutrinos \cite{Q}.

The idea of using a rolling scalar field in order to achieve a
phase of accelerated expansion in the Universe history is, of
course, not new. In fact, it is the basis of the inflationary
paradigm, where the scalar field is the so-called `inflaton'
\cite{Guth:1980zm}. Hence, quintessence corresponds to nothing
more than a late time inflationary period, taking place at
present. In this respect, the credibility of the quintessence idea
has been enhanced by the fact that the generic predictions of the
inflationary paradigm in the Early Universe are very much in
agreement with the observations.

Since they are based on the same idea, it was natural to attempt
to unify early Universe inflation with quintessence.
Quintessential inflation was thus born \cite{quinf,QI,jose,eta}.
The advantages of such unified models are many. Firstly,
quintessential inflation models allow the treatment of both
inflation and quintessence within a single theoretical framework,
with hopefully fewer and more tightly constrained parameters.
Another practical gain is the fact that quintessential inflation
dispenses with a tuning problem of quintessence models; that of
the initial conditions for the quintessence field. It is true that
attractor - tracker quintessence alleviates this tuning but the
problem never goes away completely. In contrast, in quintessential
inflation the initial conditions for the late time accelerated
expansion are fixed in a deterministic manner at the end of
inflation. Finally, a further advantage of unified models for
inflation and quintessence is the economy of avoiding to introduce
yet again another unobserved scalar field to explain the late
accelerated expansion.

For quintessential inflation to work one needs a scalar field with
a runaway potential, such that the minimum has not been reached
until today and, therefore, there is residual potential density,
which, if dominant, can cause the observed accelerated expansion.
String moduli fields appear suitable for this role because they
are typically characterised by such runaway potentials. The
problem with such fields, however, is how to stabilise them
temporarily, in order to use them as inflatons in the early
Universe.

In this paper (see also Ref.~\cite{BuenoSanchez:2006eq}) we
achieve this by considering that, during its early evolution our
modulus crosses an enhanced symmetry point (ESP) in field space.
As shown in Ref.~\cite{Kofman:2004yc}, when this occurs the
modulus may be trapped at the ESP for some time. This can lead to
a period of inflation, typically comprised by many sub-periods of
different types of inflation such as trapped, eternal, old,
slow-roll, fast-roll and so on. After the end of inflation the
modulus picks up speed again in field space resulting into a
period of kinetic density domination, called `kination'
\cite{kination}. Kination is terminated when the thermal bath of
the hot big bang (HBB) takes over. During the HBB, due to
cosmological friction \cite{cosmofric}, the modulus freezes
asymptotically at some large value and remains there until the
present, when its potential density becomes dominant and drives
the late-time accelerated expansion \cite{eta}.

Is is evident that, in order for the scalar field to play the role
of quintessence, it should not decay after the end of inflation.
Reheating, therefore should be achieved by other means. In this
paper we assume that the thermal bath of the HBB is due to the decay
of some curvaton field \cite{curv} as suggested in
Refs.~\cite{curvreh}. Note that the curvaton can be a realistic
field, already present in simple extensions of the standard model
(for example it can be a right-handed sneutrino \cite{sneu}, a flat
direction of the MSSM and its extensions \cite{mssm,nmssm} or a
pseudo Nambu-Goldstone boson \cite{pngb,orth} possibly associated
with the Peccei-Quinn symmetry \cite{PQ} etc.).  Thus, by
considering a curvaton we do not necessarily add an ad~hoc degree of
freedom. The importance of the curvaton lies also in the fact that
the energy scale of inflation can be much lower than the grand
unified scale \cite{liber}. In fact, in certain curvaton models, the
Hubble scale during inflation can be as low as the electroweak scale
\cite{orth,low}.

Throughout our paper we use natural units, where
\mbox{$c=\hbar=1$} and Newton's gravitational constant is
\mbox{$8\pi G=m_P^{-2}$}, with \mbox{$m_P=2.4\times 10^{18}$GeV}
being the reduced Planck mass.

\section{\label{sec:Fidyn}Field dynamics at an ESP}
String theory compactifications possess distinguished points in
their moduli space at which some massive states of the theory
become massless. This often results in the enhancement of the
gauge symmetries of the theory \cite{Hull:1995mz}. However, this
enhancement does not persist when the field moves away from the
point of symmetry, and therefore, such points can be associated
with symmetry breaking processes \cite{Green:1995ga}, and with a
string theoretical Higgs effect leading to moduli stabilisation
\cite{Watson:2004aq}.

Even though from the classical point of view an enhanced symmetry
point (ESP) is not a special point, as a field approaches it
certain states in the string spectrum become massless
\cite{Watson:2004aq}. In turn, these massless modes create an
interaction potential that may drive the field back to the
symmetry point. In this sense, quantum effects make the ESP a
preferred location in field space.

With respect to the properties of the scalar potential, we have that
the ESPs are dynamical attractors for the action of the fields.
Therefore the scalar potential is flat at these points
\begin{equation}\label{eq:firstderiv}
V^{\prime}_0\equiv V^{\prime}(\phi_0)=0\,,
\end{equation}
where the prime denotes derivative with respect to the modulus
$\phi$, and $\phi_0$ denotes the position of the ESP. In
Ref.~\cite{Kadota:2003tn}, the authors consider that the presence
of the ESP generates and extremum in the scalar potential. In
particular, both a maximum and a minimum are jointly considered
(requiring the presence of two ESPs).
In this paper we study the case of a single symmetry point, and
consider the cases when the ESP results in a flat inflection point
and in a local extremum.

\subsection{Particle production at an ESP}
The Lagrangian of the system for the case of two real scalar fields
$\phi$ and $\chi$ is \cite{Kofman:2004yc}
\begin{equation}\label{eq:Lagrangian}
{\cal L}=\frac{1}{2}\,\partial_{\mu}\phi\partial^{\mu}\phi
+\frac{1}{2}\,\partial_{\mu}\chi\partial^{\mu}\chi-V(\phi)-V_{\rm
int}(\phi,\chi)\,,
\end{equation}
where
\begin{equation}
  V_{\rm int}(\phi,\chi)\equiv\frac{1}{2}\,g^2\chi^2\phi^2
\end{equation}
and $g$ is a dimensionless coupling. The effective masses for the
fields $\chi$ and $\phi$ are
\begin{equation}\label{eq:effmassdef}
m^2_{\chi}=g^2\phi^2\quad,\quad m^2=m_{\phi}^2+g^2\chi^2\,,
\end{equation}
where $m_{\phi}^2\equiv V^{\prime\prime}(0)$. For simplicity, and
until Sec.~\ref{sec:Siqi}, we take $V(\phi)$ as approximately
constant and equal to \mbox{$V_0\equiv V(0)$} in the region of
interest around the ESP. Also, we write
\begin{equation}
  V_0\equiv 3H_*^2m_P^2\,,
\end{equation}
where $H_*$ is a constant mass scale equal to the Hubble scale
during inflation.

If $\phi$ moves towards the symmetry point, then $m_{\chi}^2$
changes with time. This time-dependent mass leads to the creation
of particles with momentum $k\lesssim(g\dot{\phi}_0)^{1/2}$, where
$\dot{\phi}_0$ denotes the velocity of $\phi$ in field space at
the location of the ESP $\phi_0=0$. The production takes place
when the field is within the production window
\cite{Kofman:2004yc}
\begin{equation}
|\phi|<\Delta\phi\sim(\dot{\phi}_0/g)^{1/2},.
\end{equation}
After the first production event, the expectation value
$\langle\chi^2\rangle$ is
\begin{equation}\label{eq:fluct}
\langle \chi^2\rangle\simeq\frac{1}{2\pi^2}\int_0^{\infty}
\frac{n_k^{(1)}\,k^2dk}{\sqrt{k^2+g^2\phi^2}}
\simeq\frac{n_{\chi}^{(1)}}{g|\phi|}
\end{equation}
when $\phi$ is outside the production window. The occupation
number after the first crossing is $n_k^{(1)}=\exp(-\pi
k^2/g\dot{\phi}_0)$ \cite{Felder:1999pv}. The density of $\chi$
particles grows then to $n_{\chi}^{(1)}\sim
g^{3/2}\dot{\phi}_0^{3/2}$, and the field moves in the linear
potential
\begin{equation}\label{eq:potential}
  V_{\rm
  eff}(\phi)\sim V(\phi_0)+gn_{\chi}^{(1)}|\phi|\,.
\end{equation}
The field climbs up the linear potential until it reaches a
turning point, at a distance $\Phi_1\sim
g^{-5/2}\dot{\phi}_0^{1/2}$. The field then bounces back towards
the ESP, where a new production event occurs. The interaction
potential is then reinforced by the newly created particles, and,
after leaving again the production window, the field bounces back
this time at a distance $\Phi_2<\Phi_1$. The process continues
until the turning point is at a distance from the ESP comparable
to the production window, i.e. $\Phi\sim\Delta\phi$; once this
point is reached, it can be considered that no more particles are
produced. The final number density of $\chi$ particles created
throughout the whole process is $n_{\chi}\sim
g^{-1/2}\dot{\phi}_0^{3/2}$ \cite{Kofman:2004yc}.

The picture we just described is simplistic, since once it crosses
the ESP, the field is always forced to fall back to the symmetry
point regardless of how far away is the first turning point
$\Phi_1$ from the symmetry point. However, in a more realistic
case, in order to avoid the overshoot problem we have to impose
the condition
\begin{equation}\label{eq:firstamp}
\Phi_1\sim g^{-5/2}\dot{\phi}_0^{1/2}< m_P\,,
\end{equation}
since for larger distances in field space the coupling softens and
the field, instead of falling back to the symmetry point, would
keep rolling down its potential \cite{Brustein:2002mp}.

\subsection{Post-production evolution}\label{sec:postp osc}
We take into account now the expansion of the Universe, and keep
assuming that the scalar potential $V(\phi)$ is flat enough not to
disturb the dynamics dictated by $V_{\rm int}(\phi)$.

When particle production finishes, the mass term
$m_{\chi}^2=g^2\phi^2$ no longer dominates over $k^2$ in
Eq.~(\ref{eq:fluct}). The $\chi$ particles then become
relativistic, and their energy is depleted as $a^{-4}$ by the
expansion of the Universe. Moreover, we can write
$\sqrt{k^2+g^2\phi^2}\sim k$, which means that
$\langle\chi^2\rangle$ becomes $\phi$-independent. At the end of
particle production, for which we set $a=1$, the field then
oscillates in the quadratic potential
\begin{equation}\label{eq:effquadrpot}
  V_{\rm int}(\phi)=g^2\langle\chi^2\rangle\phi^2\sim
  (g\dot{\phi}_0)\phi^2\,.
\end{equation}

Here we state the main results concerning the evolution of the
field $\phi$ and the expectation value $\langle\chi^2\rangle$,
reserving the technical details of the computation for appendix
\ref{trapdynamics}. In this appendix we find that, as long as
$m^2\gtrsim H^2$, the amplitude of oscillations $\Phi$ and the
expectation value $\langle\chi^2\rangle$ are given by
\begin{eqnarray}\label{eq:evolution1}
&\displaystyle\Phi(t)=\frac{\Phi_{a=1}}{a}\sim\frac{\Delta\phi}{a}\,,&\\
\label{eq:evolution2}
&\displaystyle\langle\overline{\chi^2}(t)\rangle=\frac{\langle\overline{\chi^2}
\rangle_{a=1}}{a^2}\sim\frac{g^{-1}\dot{\phi}_0}{a^2}=\Phi^2(t)\,,&
\end{eqnarray}
From the above, it is evident that the density of $\phi$ stored in
the oscillations around the ESP is depleted as
\begin{equation}\label{eq:oscil rate}
  \rho_{\rm osc}=V_{\rm int}(\Phi)\propto a^{-4}\,.
\end{equation}
We also stress that as long as the oscillation energy dominates
the total density, the ratio $m^2/H^2$, where $m^2\approx
g^2\langle\overline{\chi^2}\rangle$, grows as the Universe
expands, $m^2/H^2\propto \langle\overline{\chi^2}\rangle/\rho_{\rm
osc}\propto a^2$.

When the kinetic density of the oscillating field falls below the
potential density $V_0$, a phase of inflation called \textit{trapped
inflation} begins \cite{Kofman:2004yc}. Taking into account the
depletion rates given by Eqs.~(\ref{eq:evolution1}) and
(\ref{eq:oscil rate}), trapped inflation commences at
$a\sim(\dot{\phi}_0^2/V_0)^{1/4}$. Using Eq.~(\ref{eq:evolution1}),
the amplitude of oscillation is
\begin{equation}
  \Phi_{i}\sim (g^{-1}H_*m_P)^{1/2}\,.
\end{equation}
From this moment on $H$ becomes approximately equal to $H_*$, and
the field starts to decrease its rate of oscillation $m^2/H^2\sim
g^2\langle\overline{\chi^2}\rangle/H_*^2\propto a^{-2}$. Owing to
this decrease, the phase of oscillations will finish when the
expansion of the Universe starts to overdamp the oscillations of
the field, which happens when $m^2/H_*^2\sim 1$. The field then
reaches its minimum amplitude of oscillations. Using
Eq.~(\ref{eq:evolution2}), this is given by
\begin{equation}
  \Phi_{\rm min}\sim
H_*/g\,.
\end{equation}
The amount of inflation during the phase of oscillations is then
\begin{equation}\label{Nosc}
  N_{\rm osc}=\ln\frac{\Phi_i}{\Phi_{\rm min}}\sim
  \ln \left[g^{1/2}\left(\frac{m_P}{H_*}\right)^{1/2}\right]\,.
\end{equation}

Note that $N_{\rm osc}$ is independent of the initial value
$\dot{\phi}_0$. This is because the kinetic density in excess of
$V_0$ must be exhausted before trapped inflation can begin.
Therefore, we find the same amount of inflation if we take
$\dot{\phi}_0^2\sim V_0$ and set the beginning of trapped
inflation at $a=1$, which corresponds to suppressing the
interphase in which the excess of kinetic density is depleted.
Hence, the only scale relevant here is $H_*$.

Once the phase of oscillation is over, the depletion rates
computed in Eqs.~$(\ref{eq:evolution1})$ and
$(\ref{eq:evolution2})$ no longer apply. The expansion of the
Universe now makes the mass of the field fall below $H_*$, and a
phase of slow-roll inflation follows. We emphasize that this phase
of slow-roll occurs with the field rolling down the interaction
potential $V_{\rm int}(\phi)$. When the expansion of the Universe
has redshifted sufficiently the kinetic density of the field, the
motion of the later ceases to be classical to become dominated by
the vacuum fluctuation of the field. A phase of eternal inflation
then follows. We devote the rest of this section to study in
detail this transient to the phase of eternal inflation.

\subsection{Towards eternal inflation}\label{sec:toetinf}
When the mass of the field falls below $H_*$ after the end of the
oscillatory phase, the field begins to slow-roll over the
interaction potential. The equation of motion is
\begin{equation}\label{eq:slow-roll}
3H\dot{\phi}\simeq-V_{\rm int}^{\prime}(\phi)\,,
\end{equation}
and the classical motion of the field $\Delta\phi_c$ per Hubble time
is
\begin{equation}\label{slowrollmot}
  \Delta\phi_{\rm
c}\sim \dot{\phi}/H_*\simeq-\frac{V_{\rm int}^{\prime}(\Phi_{\rm
min})}{3H_*^2}\,,
\end{equation}
Using the estimate above we can compare the classical motion
$\Delta\phi_c$ with $|\phi|\lesssim\Phi_{\rm min}$ at the
beginning of slow-roll, obtaining
\begin{equation}
  \frac{|\Delta\phi_c|}{\Phi_{\rm min}}
  \sim\frac{|V_{\rm int}^{\prime}|}{H_*^2}\frac{1}{\Phi_{\rm
  min}}\sim
  \frac{g^2\langle\chi^2\rangle|\phi|}{H_*^2\,\Phi_{\rm min}}<1\,,
\end{equation}
which follows from the slow-roll condition $|\eta|<1$. Therefore,
the value $|\phi|$ after the end of oscillations remains within
the order of $\Phi_{\rm min}$, and roughly constant during a
Hubble time $|\phi|\lesssim \Phi_{\rm min}$. Moreover, given that
both the kinetic term $k^2$ and the mass term $g^2\Phi^2$ in
$\langle\overline{\chi^2}\rangle$ [c.f. Eq.~(\ref{eq:fluct})]
remain comparable until the end of oscillations, the mass term
starts to dominate over the kinetic term soon after the onset of
the slow-roll phase. Consequently, the $\chi$ particles become
again non-relativistic. Indeed,
\begin{equation}\label{eq:x2rate}
  \langle\chi^2\rangle\sim\int_0^{\infty}
\frac{n_k(t)\,k^2dk}{\sqrt{k^2+g^2\phi^2}}\
  \propto\frac{n_{\chi}}{|\phi|}\propto a^{-3}\,.
\end{equation}
This result allows us to obtain the decrease rate of both the slope
$V^{\prime}_{\rm int}(\phi)\propto a^{-3}$ and the effective mass
squared $m^2\propto a^{-3}$. The scaling law for the kinetic density
of the field
\begin{equation}\label{kinslowroll}
\rho_{\rm kin}=\frac{1}{2}\,\dot{\phi}^2\simeq\frac{1}{2}
\left(\frac{V_{\rm int}^{\prime}}{3H}\right)^2\propto a^{-6}\,,
\end{equation}
which may be compared to the milder rate that applies when $\phi$
oscillates, Eq.~(\ref{eq:oscil rate}).

Now we compute the amount of slow-roll inflation during the
transient to eternal inflation. Owing to rapid redshift of the
effective mass and to the very slow-roll motion of the field, we
assume that $|\phi|$ remains within the order of $\Phi_{\rm min}$.

The slow-roll motion of the field begins when $m^2\sim H_*$, and
persists until
\begin{equation}\label{eq:cmend}
|V^{\prime}_{\rm int}|\sim H_*^3\,.
\end{equation}
Using that $|V^{\prime}_{\rm int}|\sim H_*^3/g$ at the beginning
of the slow-roll and that $V^{\prime}_{\rm int}\propto a^{-3}$,
the amount of slow-roll inflation during the transient to eternal
inflation is
\begin{equation}\label{numsr}
  N_{\rm sr}\sim\frac{1}{3}\ln \frac{1}{g}\,.
\end{equation}
Taking for example $g\simeq 0.1$ this gives $N_{\rm sr}\sim 1$,
which is a negligible amount. The assumption $|\phi|\sim\Phi_{\rm
min}$ is indeed correct.

Once $|V^{\prime}_{\rm int}|$ falls below $H_*^3$, the field becomes
dominated by its vacuum fluctuation. The field then performs a
random motion taking steps of amplitude $\delta\phi\sim H_*/2\pi$
every Hubble time. The motion of the field during this phase is
oblivious of the scalar potential and is only known
probabilistically\footnote{Strictly speaking, the motion of the
field becomes unaffected by the scalar potential when
\mbox{$\bar{\phi}^2\lesssim H_*^4/m^2$} \cite{Starobinsky:1994bd}.
When Eq.~(\ref{eq:cmend}) is satisfied, the field finds itself
outside this region. Owing though to $m^2\propto a^{-3}$, this
region grows exponentially with time, and very soon the field finds
itself within it. This intermediate regime does not result in any
significant difference with respect to the cases we address in
Sec.~\ref{sec:infl dyn}, and so we omit it for simplicity.}, with
the root mean square of the field growing to
$\sqrt{\langle\phi^2\rangle}\sim \frac{H_*}{2\pi}\sqrt{N}$ after $N$
$e$-foldings of eternal inflation \cite{Vilenkin:1982wt}. The number
of $e$-foldings of trapped inflation is therefore well approximated
by Eq.~(\ref{Nosc}). In view of this result we emphasize that, as
far as trapped inflation in concerned, \emph{the main virtue of the
trapping mechanism does not rely on its capacity to produce
inflation, but in that it sets the field at a locally flat region in
field space}. Therefore, depending upon the strength of the symmetry
point, this mechanism may fix the initial conditions leading to a
subsequent long-lasting stage of inflation.

Up until now we have assumed that the scalar potential $V(\phi)$
plays no role in the dynamics of the field, apart from
contributing $V_0$. We now address the complete picture where a
non-trivial `background' scalar potential $V(\phi)$ is present.

\section{String inspired quintessential inflation}\label{sec:Siqi}
Type IIB compactifications have received a great deal of attention
due to recent accomplishments in moduli stabilisation
\cite{Giddings:2001yu}. The integrated contribution of the fluxes
induces a non-zero superpotential, $W=W_0$. Assuming that this
contribution is independent of the fields leads to no-scale
supergravity, in which the K\"ahler moduli remain exactly flat.
However, these flat directions can be lifted up through
non-perturbative effects, such as gaugino condensation or D-brane
instantons \cite{Derendinger:1985kk}. In this case, the
superpotential $W$ takes the form
\begin{equation}\label{W}
W=W_0+W_{\rm np}\quad \textrm{with} \quad W_{\rm np}=Ae^{-cT}\,,
\end{equation}
where $W_0$ is the tree level contribution from fluxes, $A$ and $c$
are constants, whose magnitude and physical interpretation depends
on the origin of the non-perturbative term (in the case of gaugino
condensation $c$ is expected to take on values $c\lesssim1$), and
$T$ is a K\"ahler modulus in units of $m_P$, for which
\begin{equation}
  T=\sigma+i\alpha\,.
\end{equation}
with $\sigma$ and $\alpha$ real. The inclusion of the
non-perturbative superpotential $W_{\rm np}$ results in a runaway
scalar potential characteristic of string compactifications. In
type IIB compactifications with a single K\"ahler modulus, this is
the so-called volume modulus. In this case, the runaway behaviour
leads to decompactification of the internal manifold.

The tree level K\"ahler potential for the a modulus, written in
units of $m_P^2$, is
\begin{equation}\label{eq:treeKahler}
  K=-3\,\textrm{ln}\,(T+\bar{T})\,,
\end{equation}
and the corresponding supergravity potential is
\begin{equation}\label{scalarpot0}
  V_{\rm np}(\sigma)=\frac{cAe^{-c\sigma}}{2\sigma^2m_P^2}\left[
  \left(1+\frac{c\sigma}{3}\right)Ae^{-c\sigma}+W_0
  \cos(c\alpha)\right]\,.
\end{equation}
To secure the validity of the supergravity approximation we consider
$\sigma>1$. For values of $c\lesssim1$, we then have $c\sigma>1$. To
account now for the ESP, we consider that the scalar potential
$V_{\rm np}(\phi)$ receives a phenomenological contribution $V_{\rm
ph}(\sigma)$ important only in the neighbourhood of the ESP.
Therefore, we consider the scalar potential
\begin{equation}\label{eq:newscalarpot}
  V(\sigma)=V_{\rm np}(\sigma)+V_{\rm ph}(\sigma)
\end{equation}
as long as the field is close to the ESP, whereas we consider
$V(\sigma)\approx V_{\rm np}(\sigma)$ when the field is
sufficiently away from it. By choosing $V_{\rm ph}(\sigma)$
appropriately it is possible to account for an ESP with
$dV/d\sigma=0$ at $\sigma_0$. The field may thus drive a period of
inflation when close to $\sigma_0$. In order to study the arising
inflationary dynamics, we turn to the canonically normalized field
$\phi$ associated to $\sigma$. Owing to Eq.~(\ref{eq:treeKahler})
we have
\begin{equation}\label{canonfield}
  \sigma(\phi)=\textrm{exp}\,(\sqrt{2/3}\,\phi/m_P)\,.
\end{equation}

\begin{figure}[htbp]
\hspace{5cm}\epsfig{file=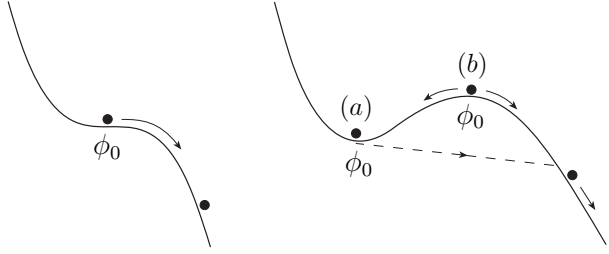, width=8cm}
\caption{\label{fig:ESPs}\small The left panel shows the scalar
potential $V(\phi)$ around an enhanced symmetry point (ESP) at
$\phi_0$ and with a subdominant quadratic term. The right panel
shows the formation of either a minimum, case (a), or a maximum,
case (b). If the ESP results in a minimum the field accesses the
large field region through quantum tunneling. We assume that after
tunneling the field appears at approximately the same height. If the
ESP results in a maximum, the field may fall towards the minimum or
continue rolling towards large values.}
\end{figure}

In appendix \ref{app:minapp} we show how, within our
phenomenological approach, the inflationary dynamics may be fully
accounted for through the scalar potential
\begin{equation}\label{eq:scpot}
  V(\phi)=V_0+\frac{1}{2}m_{\phi}^2\bar{\phi}^2+\frac{1}{3!}
  \vppp\bar{\phi}^3\,,
\end{equation}
where we have defined
\begin{equation}
  \bar{\phi}\equiv\phi-\phi_0\,.
\end{equation}
Also, with the symmetry point located at $\phi_0$, the interaction
potential becomes now \mbox{$V_{\rm
int}(\phi)=\frac{1}{2}g^2\chi^2\bar{\phi}^2$}. The free parameters
of the scalar potential $V(\phi)$ are $H_*$, $m_{\phi}^2$ and
$\vppp$. Within our minimal approach, $\vppp$ is determined by
$H_*$, which sets the height of the potential at the ESP, and by how
close to the symmetry point the potential $V_{\rm ph}(\sigma)$
becomes exponentially suppressed with respect to $V_{\rm
np}(\sigma)$ (see Appendix \ref{app:minapp}). In our model we allow
these two parameters to vary independently. Also, the sign of
$\vppp$ is always negative, whereas $m_{\phi}^2$ may be positive,
negative, or effectively zero if the quadratic term in $V(\phi)$ is
subdominant with respect to the cubic one. These three cases are
pictured in Fig \ref{fig:ESPs}.

If the ESP results in a maximum, the field may grow to large values
when the field breaks free from its trapping even if the quadratic
term dominates $V(\phi)$. Nonetheless, if the quadratic term
dominates there is still a chance that the field is stabilised in
the neighbouring minimum. On the contrary, an ESP resulting in a
minimum always leads to stabilisation if the quadratic term
dominates $V(\phi)$ when the field breaks free. In such a case the
field drives a period of \textit{old} inflation \cite{Guth:1980zm}.
Once the field tunnels the barrier, it continues rolling down its
potential.

Stabilisation when the ESP results in a minimum is considered in
Ref.~\cite{Dine:1998qr}, where the suggestion is made of stabilising
all of the moduli fields at points of enhanced symmetry (see also
Ref.~\cite{Greene:2007sa} for a more recent study of stabilisation
at an ESP). In contrast to these cases, the ESP considered in our
approach just provides an effective ground state unstable through
tunneling.

\section{Inflationary dynamics}\label{sec:infl dyn}

\subsection{Inflationary regimes}\label{sec:ShapeESP}
In Sec.~\ref{sec:Fidyn} we assumed that $V(\phi)$ is flat enough
to allow the field to progress ultimately to the phase of eternal
inflation under only the influence of the interaction potential
$V_{\rm int}$. In general, however, the scalar potential $V(\phi)$
may start to drive the dynamics at any moment after the end of
particle production. This happens when
\begin{equation}\label{eq:slopes}
  |V^{\prime}(\Phi)|\sim V_{\rm int}^{\prime}(\Phi)\,.
\end{equation}
From that moment onwards the field breaks free from its trapping and
continues rolling over the scalar potential $V(\phi)$. The
subsequent evolution of the field may be readily classified in terms
of the quantity
\begin{equation}\label{eq:f(g)}
  f(g)\equiv\left|g\frac{m_{\phi}^2}{3H_*^2}+\frac{1}{2}\frac{\vppp}
{\sqrt{3}H_*}\right|\,.
\end{equation}
We find the following cases:

\noindent \textbf{Case 1:} The field breaks free during the period
of fast oscillations following particle production. This corresponds
to $|V^{\prime}(\Phi_{\rm min})|>V^{\prime}_{\rm int}(\Phi_{\rm
min})$, i.e. $V(\phi)$ controls the dynamics of the field before the
latter ceases to oscillate at $|\bar{\phi}|\sim\Phi_{\rm min}$.
Using Eq.~(\ref{eq:scpot}), this condition translates into
\begin{equation}
  g<f(g)\,.
\end{equation}
In this case, the field is unable to drive a sufficiently
long-lasting period of inflation.

\noindent \textbf{Case 2:} The field escapes from its trapping at
the end of the oscillations. This corresponds to
$|V^{\prime}(\Phi_{\rm min})|\sim V^{\prime}_{\rm int}(\Phi_{\rm
min})$, which translantes into
\begin{equation}
  f(g)\sim g\,.
\end{equation}
If $V^{\prime\prime}(\phi)$ does not increase significantly away
from the ESP, the field may drive a limited period of inflation,
known as \emph{fast-roll} inflation \cite{Linde:2001ae}.

\noindent \textbf{Case 3:} The field breaks free during the short
phase of slow-roll following the end of oscillations. In order to
avoid that the field becomes driven by its vacuum fluctuation, we
must impose the lower bound $H_*^3\lesssim V^{\prime}(\Phi_{\rm
min})$. This case thus corresponds to
\begin{equation}\label{eq:intsecsr2}
    g^2\lesssim f(g)< g\,.
\end{equation}
The slow-roll parameters when the field starts to roll over the
scalar potential $V(\phi)$ are $\varepsilon\ll1$ and $\eta<1$, and
hence the field continues its slow-roll motion when released from
its trapping. This case may result in enough slow-roll inflation
to solve the flatness and horizon problems.

\noindent \textbf{Case 4:} The motion of the field is dominated by
its vacuum fluctuation. In this case, the field goes through the
short phase of slow-roll and then starts to drive eternal inflation.
From this stage onwards, the interaction potential becomes very
redshifted, playing no role in the dynamics of the field any longer.
The field then finds itself at a very flat patch of the scalar
potential $V(\phi)$ and driven by its vacuum fluctuation. The field
performs a random motion taking steps of amplitude $\delta\phi\sim
H_*$, and the root mean square of the displaced field grows as
$\sqrt{\langle\bar{\phi}^2\rangle}\sim NH_*$ \cite{Vilenkin:1982wt}.
This case happens for
\begin{equation}\label{eq:inteternal}
    f(g)< g^2\,.
\end{equation}

With time, in some parts of the Universe the displaced field grows
large, and the steepness of the scalar potential $V(\phi)$
increases. When $V(\phi)$ becomes steep enough the field recovers
its classical motion. Eternal inflation then finishes and a
long-lasting phase of slow-roll inflation follows. This final phase
of slow-roll inflation may be long enough as to solve the flatness
and horizon problems as well.

In other parts of the Universe, though, the displaced field always
remains in the very flat region. The field never recovers its
classical motion, and so it continues driving a never-ending phase
of inflation.

\subsection{Parameter space for observable inflation}
We focus first on the inflationary regimes able to produce enough
slow-roll inflation to solve the flatness and horizon problems. In
general, the inflationary regimes able to do that provide a number
of slow-roll $e$-foldings far larger than the required by
observation. However, we only focus on the observable amount of
inflation. Also, we allow the observed value of the curvature
perturbation to be generated, either partially or entirely, by a
scalar field other than the inflaton. The only condition to allow
this is that the curvature perturbation generated by the inflaton
field, which we denote by $\pphi$, is not excessive. The CMB
normalization of the spectrum of inflaton perturbations:
$\pphi\simeq5\times10^{-5}$ \cite{Liddle:2000cg}, then becomes a
bound \cite{liber}
\begin{equation}\label{eq:CMBbnd}
  \pphi\lesssim10^{-5}\,.
\end{equation}

As explained in Sec.~\ref{sec:Siqi} , the inflationary dynamics in
our phenomenological approach can be fully described through the
scalar potential in Eq.~(\ref{eq:scpot}). In this kind of potential,
the number of $e$-foldings may be well approximated by taking the
end of inflation in the limit $\bar{\phi}(|\eta|\sim1)\to\infty$.
Using the slow-roll formula \cite{Liddle:2000cg} in this limit, we
obtain $\bar{\phi}$ when cosmological scales leave the horizon $N$
$e$-foldings before the end of inflation\footnote{For the derivation
of this formula in the case of a similar potential to
Eq.~(\ref{eq:scpot}) see \cite{BuenoSanchez:2006xk}.}
\begin{equation}\label{eq:phibar}
  \bar{\phi}=\frac{m^2}{|V_0^{\textrm{\tiny
(3)}}|}+\frac{|m^2|}{|V_0^{\textrm{\tiny (3)}}|}\coth(N|\eta|/2)\,.
\end{equation}
The curvature perturbation generated by the inflaton field is
\begin{equation}\label{eq:pert}
  {\cal P}^{1/2}_{\phi}=\frac{1}{2}\frac{N^2|\vppp|}{\sqrt{3}H_*}
  \left(\frac{\sinh (N|\eta|/2)}{N|\eta|/2}\right)^2\approx
  \frac{1}{2}\frac{N^2|\vppp|}{\sqrt{3}H_*}\,,
\end{equation}
where the last follows from the slow-roll condition $|\eta|<1$, and
typical values are $N=45-60$. This equation determines $\vppp$ in
terms of $H_*$ and $\pphi$. The advantage of this change, apart from
the clearer physical meaning of $\pphi$, is that
Eq.~(\ref{eq:CMBbnd}) allows us to pick out easily the parameter
space leading to an acceptable inflationary cosmology. Also, because
in our phenomenological approach $\vppp$ has a lower bound at fixed
$H_*$ (see Appendix \ref{app:minapp} ), so has $\pphi$. Therefore,
$\pphi$ takes on values in the interval
\begin{equation}\label{eq:intpert}
  N^2(c\sigma_0)^3\frac{H_*}{m_P}<\pphi\lesssim10^{-5}\,.
\end{equation}

Consistency of this inequality sets a maximum value for $H_*$,
thus resulting in an upper bound for the inflationary scale
\begin{equation}\label{eq:upperm0}
  V_0^{1/4}\lesssim10^{12}\textrm{ GeV}\,,
\end{equation}
where we have taken $c\sigma_0\sim10$ and $N=60$. The bound is
saturated when the curvature perturbation is generated by the
inflaton.

Now, in order to determine what inflationary regimes can be realised
in our model, we must find out if they are compatible with the field
being trapped. Note that in all the previous discussion to identify
the different inflationary regimes, we have implicitly assumed that
the field is trapped in the interaction potential $V_{\rm
int}(\phi)$. Nonetheless, this might not be the case, hence it is
then crucial to determine the interval of $g$ leading to trapping.

To secure that the field is trapped after crossing the ESP, it
suffices to require that the scalar potential $V(\phi)$, apart
from contributing $V_0$, plays no role in the dynamics during
particle production. Owing to the decreasing amplitude of
oscillations during this process, it is enough to demand
$|V^{\prime}(\Phi_1)|<|V_{\rm int}^{\prime}(\Phi_1)|$, where
$\Phi_1$ is the first turning point of the oscillation (see
Eq.~(\ref{eq:firstamp})). This condition translates into
\begin{equation}
g^{10}>\frac{m_{\phi}^2}{m_P^2}+\frac{H_*}{m_P}N^{-2}\pphi\,,
\end{equation}
where we have used $|V^{\prime}_{\rm int}(\Phi_1)|\sim
gn_{\chi}^{(1)}$ [c.f. Eq.~(\ref{eq:potential})], and
Eq.~(\ref{eq:pert}). Using now the lower bound in
Eq.~(\ref{eq:intpert}) and the slow-roll condition $|\eta|<1$
(which requires $m_{\phi}^2<H_*^2$), the first term on the
right-hand side becomes negligible. The values of $g$ leading to
trapping are thus given by
\begin{equation}\label{eq:trp}
g^{10}\left(\frac{m_P}{H_*}\right)>N^{-2}\pphi\,.
\end{equation}
Taking for example the upper bound found in Eq.~(\ref{eq:upperm0})
and $\pphi\sim10^{-5}$, we obtain $g>10^{-2}$. It is also clear
that if $\pphi$ and $H_*$ decrease, so does the lower bound for
$g$, thus widening the range of values leading to trapping. Using
for example the lower bound in Eq.~(\ref{eq:intpert}) we obtain
$g>(H_*/m_P)^{1/5}$.

\begin{figure}[htbp]
\hspace{1cm}\epsfig{file=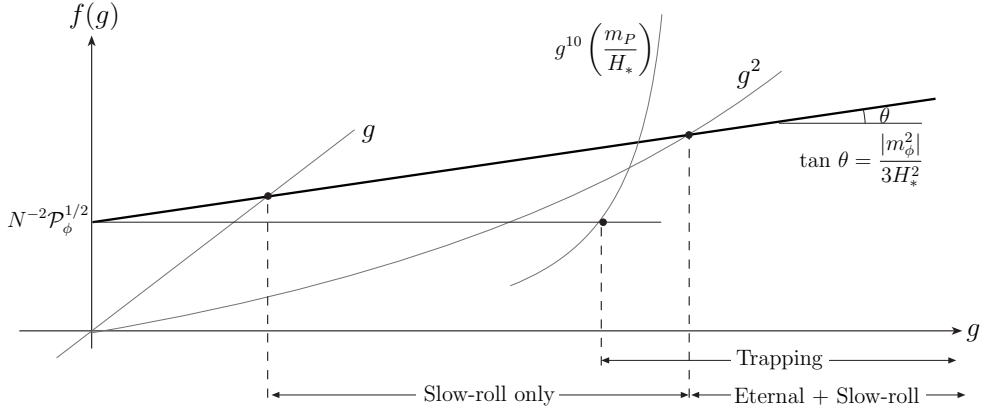, width=13cm}
\caption{\label{fig:infmap1}\small The figure shows the intervals
in the $g$-axis corresponding to the two inflationary regimes in
terms of the parameters $\pphi,m_{\phi}^2$ and $H_*$ for
$m^2_{\phi}\leq0$. For given values of the parameters, an
inflationary regime occurs only occur if its interval in the
$g$-axis has significant overlapping with the interval
corresponding to trapping. The solid line represents the function
$f(g)$, Eq.~(\ref{eq:f(g)}), whose intersections with the curves
$g$ and $g^2$ determine the different inflationary regimes, [c.f.
Eqs.~(\ref{eq:intsecsr2}) and (\ref{eq:inteternal})]. The
intersection of the horizontal line $N^{-2}\pphi$ with the curve
$g^{10}(m_P/H_*)$ determine the interval for trapping [c.f.
Eq.~(\ref{eq:trp})]}
\end{figure}

Let us study first the case $m_{\phi}^2\leq0$, i.e. the ESP results
in either a maximum or a flat inflection point. The results are
pictured in Figure \ref{fig:infmap1}. %
Solving for $g$ in Eq.~(\ref{eq:intsecsr2}) and using
Eq.~(\ref{eq:pert}), we obtain the interval in which only slow-roll
inflation occurs:
\begin{equation}\label{eq:onlySR}
  N^{-2}\pphi<
g\lesssim\frac{|m_{\phi}^2|}{6H_*^2}+\sqrt{\left(\frac{|m_{\phi}^2|}{6H_*^2}\right)^2
+N^{-2}\pphi},
\end{equation}
whereas solving for $g$ in Eq.~(\ref{eq:inteternal}) we find the
interval where eternal inflation plus slow-roll inflation occurs:
\begin{equation}
  \frac{|m_{\phi}^2|}{6H_*^2}+\sqrt{\left(\frac{|m_{\phi}^2|}{6H_*^2}\right)^2
+N^{-2}\pphi}<g\lesssim1\,.
\end{equation}

If the cubic term dominates at $\bar{\phi}\sim\Phi_{\rm min}$ the
ESP may be effectively considered as a flat inflection point, and
we can take $m_{\phi}^2\approx0$. From Fig.~\ref{fig:infmap1} it
is clear that the inflationary regime with eternal inflation plus
slow-roll inflation is always compatible with trapping. However,
the inflationary regime with only slow-roll inflation is
compatible with trapping if the upper bound in
Eq.~(\ref{eq:onlySR}) is at least of the order of the lower bound
in Eq.~(\ref{eq:trp}). This in turn requires far too low an
inflationary scale, $V_0^{1/4}\lesssim10^2$ GeV. The only
inflationary regime able to solve the flatness and horizon problem
thus involves a phase of eternal inflation. This result is
illustrated in Fig.~\ref{fig:infmap1}, where the intervals
corresponding to only slow-roll inflation and trapping have no
overlapping in the limit $m_{\phi}^2\approx0$.

Assume now that the quadratic term dominates at
$\bar{\phi}\sim\Phi_{\rm min}$. The upper bound in
Eq~(\ref{eq:onlySR}) is now $g\sim|m_{\phi}^2|/H_*^2$. Inflation
with only slow-roll is compatible with trapping if this value of
$g$ is at least of the order of the lower bound in
Eq.~(\ref{eq:trp}), i.e.
\begin{equation}
|\eta|\gtrsim\left(N^{-2}\pphi\frac{H_*}{m_P}\right)^{1/10}\,.
\end{equation}

If the inflaton contributes substantially to the curvature
perturbation, i.e. $\pphi\sim10^{-5}$, a sufficiently high
inflationary scale is only achieved if $|\eta|$ satisfies the bound
\begin{equation}\label{eq:eta}
  |\eta|\geq10^{-2}\,.
\end{equation}
However, using the lower bound for $\pphi$ in
Eq.~(\ref{eq:intpert}) allows for smaller values
$|\eta|\geq10^{-3}$. We thus conclude that the inflationary regime
where only slow-roll occurs is compatible with trapping if the
quadratic term dominates $V(\phi)$ when the field starts to roll
over it. This result is also illustrated in
Fig.~\ref{fig:infmap1}, where the regime with only slow-roll has
overlapping with the trapping interval for sufficiently large
values of $g\sim|m_{\phi}^2|/H_*^2$.

Consider now $m_{\phi}^2>0$. This case is different from the last
one because now $m_{\phi}^2$ and $\vppp$ have different sign.
Therefore, $f(g)$ may feature two branches. The decreasing branch
always intersects the curve $g^2$
\begin{equation}\label{eq:g1}
  g\sim-\frac{m_{\phi}^2}{6H_*^2}+\sqrt{\left(\frac{m_{\phi}^2}
  {6H_*^2}\right)^2+N^{-2}\pphi}\,,
\end{equation}
However, the growing branch (if present for $g\lesssim1$) may or
may not intersect the curve $g^2$ (see Fig.~\ref{fig:infmap2}).
The intersection points of the growing branch with $g^2$ are
\begin{equation}\label{eq:g2}
  g_{\pm}\sim\frac{m_{\phi}^2}{6H_*^2}\pm\sqrt{\left(\frac{m_{\phi}^2}
  {6H_*^2}\right)^2-N^{-2}\pphi}\,,
\end{equation}
which do not exist unless the mass term dominates over the cubic
one. If that is the case, $g$ given in Eq.~(\ref{eq:g1}) and $g_-$
in Eq.~(\ref{eq:g2}) converge to the same value. Also, because of
the slow-roll condition, $f(g)$ intersects only once with the
straight line $g$ at $g\sim N^{-2}\pphi$. As a result, if the
quadratic term dominates $V(\phi)$ at $\bar{\phi}\sim\Phi_{\rm
min}$, the interval in which only slow-roll inflation occurs is
\begin{equation}
  N^{-2}\pphi<g\lesssim \frac{m_{\phi}^2}{3H_*^2}\,,
\end{equation}
which coincides with the result in Eq.~(\ref{eq:onlySR}) in the
same limit.

Finally, it is obvious that both results coincide in the limit
$m_{\phi}^2\approx0$. The results found are then the same: %
\begin{figure}[htbp]
\hspace{1cm}\epsfig{file=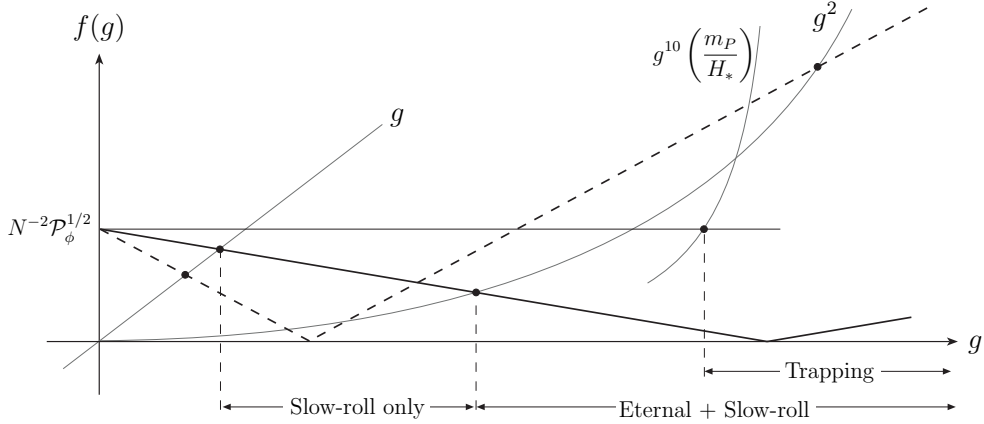, width=13cm}
\caption{\label{fig:infmap2}\small Intervals in the $g$-axis
corresponding to the two inflationary regimes for
$m_{\phi}^2\geq0$. The function $f(g)$ features now a decreasing
branch owing to $\vppp$ and $m^2$ having different sign. If there
is a growing branch, this may or may not intersect the curve
$g^2$, depending on which term dominates $V(\phi)$ at
$\bar{\phi}\sim\Phi_{\rm min}$. The ESP may be effectively
considered as having $m_{\phi}^2\approx0$ if the curve $g^2$ is
intersected only once (bold line). Otherwise the mass term
dominates the scalar potential and the field is temporarily
stabilised at the minimum (dashed line).}
\end{figure}
If the quadratic term dominates at $\bar{\phi}\sim\Phi_{\rm min}$,
the growing branch has two intersections with the curve $g^2$,
allowing the regime with slow-roll only to be compatible with
trapping for sufficiently large values of $m_{\phi}^2/H_*^2$. If
the quadratic term is subdominant, the growing branch has no
intersections with $g^2$, and the only intersection is in the
decreasing branch. In this case, the limit $m_{\phi}^2\approx0$ is
recovered. These two possibilities are illustrated in
Fig.~\ref{fig:infmap2}.

Up to now, we have used a phase of slow-roll inflation to solve
the flatness and horizon problems. However, it is also possible to
solve these problems without slow-roll inflation, but with
$|\eta|\sim1$. The resulting phase of inflation is called
\textit{fast roll} inflation \cite{Linde:2001ae}. In our model,
this possibility may arise if the ESP results in a maximum with
$|m_{\phi}^2|\sim H_*^2$. From Eq.~(\ref{eq:scpot}), the effective
mass of the field is $m^2=m_{\phi}^2+\vppp\bar{\phi}$.  Therefore,
$m^2$ remains within the order of $H_*^2$ for field displacements
of $\bar{\phi}\lesssim m_{\phi}^2/\vppp$. Fast-roll inflation then
finishes at $\bar{\phi}_{\rm e}\sim m_{\phi}^2/\vppp$, since for
larger values $|m^2|>H_*^2$.

If the ESP results in a minimum where the field is temporarily
stabilised. After tunneling the potential barrier, the field finds
itself at distances of the order $m_{\phi}^2/|\vppp|$. From that
point onwards, enough inflation is only possible if $|\eta|<1$,
which corresponds to slow-roll inflation. Fast-roll inflation is
insufficient in this case.

Assuming then that the ESP results in a maximum, the equation of
motion for $\phi$ may be approximated by
\begin{equation}
  \ddot{\bar{\phi}}+3H\dot{\bar{\phi}}-|m_{\phi}^2|\bar{\phi}\simeq0
\end{equation}
for field values $\bar{\phi}\lesssim\bar{\phi}_{\rm e}$. This
equation may then be solved \cite{Linde:2001ae}, and the field value
$N$ $e$-foldings before the end of fast-roll inflation given
\begin{equation}\label{eq:fastroll}
  \bar{\phi}\sim\bar{\phi}_{\rm e}\exp(-NF_{\phi})\,,
\end{equation}
where $F_{\phi}$ is
\begin{equation}\label{eq:Ffastroll}
  F_{\phi}\equiv\frac{3}{2}\left(\sqrt{1+\frac{4}{3}|\eta|}-1\right)\,.
\end{equation}

The amplitude of perturbation generated by the inflaton when the
observable Universe leaves the horizon is $\pphi\sim
H_*\delta\phi/\dot{\phi}$ \cite{Liddle:2000cg}. Writing
$N=H_*\Delta t$ we obtain $\dot{\phi}$ from
Eq.~(\ref{eq:fastroll}), and taking $\delta\phi\sim H_*$ we have
\begin{equation}
  \pphi\sim\frac{H_*\exp(NF_{\phi})}{F_{\phi}}\frac{\vppp}{m_{\phi}^2}\,.
\end{equation}
Using the lower bound for $|\vppp|$, Eq.~(\ref{eq:thirdderiv3}),
we obtain
\begin{equation}\label{eq:Hfr}
  \frac{H_*}{m_P}<\frac{|\eta|F_{\phi}}{\exp(NF_{\phi})}
  \frac{\pphi}{(c\sigma_0)^3}\,,
\end{equation}
thus constraining $H_*$ in terms of $|\eta|$. The bound is less
stringent when the inflaton contributes significantly to the
curvature perturbation, in which case
\begin{equation}
H_*<10^6\frac{|\eta|F_{\phi}}{\exp(NF_{\phi})}\,\,\textrm{TeV}\,,
\end{equation}
where we take $c\sigma_0\sim10$.

\section{Quintessence}\label{sec:RaQ}
\subsection{After the end of inflation}
In the first stage of its evolution after inflation, the energy
density of the field becomes again dominated by its kinetic energy
because of the steepness of the scalar potential $V_{\rm np}$.
This phase is known as kination, and the evolution of the field is
governed by the equation
\begin{equation}\label{eqnkination}
\ddot{\phi}+3H\dot{\phi}\simeq0\,.
\end{equation}

In order to achieve a successful model for quintessential
inflation, the field cannot decay after the end of inflation, for
otherwise it could not have survived until today to become
quintessence. The usual assumption to get around this is to
consider that the couplings of $\phi$ to the standard model
particles are suppressed so that no instant preheating mechanism
applies \cite{Felder:1999pv}, thereby avoiding the decay of the
field into a thermal bath of the standard model particles. This is
reasonable to assume for a modulus field when away from ESPs.

To recover the Hot Big Bang (HBB) it is necessary to discuss an
alternative method to achieve reheating after inflation. One
possibility to do this is through the decay of a curvaton field
\cite{curvreh}. As an additional advantage, this curvaton field
helps to produce the correct amplitude of the primordial density
perturbations \cite{curv}, while it also allows more easily
inflation to occur at relatively low energies \cite{liber,low}.

In this case, the phase of kination finishes when the curvaton, or
its decay products, catches up with the kinetic energy of the field,
eventually leading to the onset of the HBB. It is possible here to
envisage two scenarios to end kination. One is to consider that the
curvaton decays after the end of kination, and therefore a period of
matter domination follows kination, during which the Universe is
dominated by the oscillating curvaton field \cite{curvreh}. The
other possibility is to consider that the curvaton field decays
before kination finishes, and the Universe becomes radiation
dominated at the end of kination. For our purposes these two cases
are roughly equivalent, for, once kination finishes, the equation of
motion of $\phi$ features an asymptotic value $\phi_F$, which does
not differ much from one case to the other. Consequently we adopt
the simplest scenario in which the curvaton field decays before it
dominates, leading to radiation domination after
kination\footnote{This is also in accordance with the fact that in
quintessential inflation, kination ends not much earlier than
nucleosynthesis, while the curvaton must decay much sooner for
baryogenesis considerations.}. Therefore, at the end of kination the
Hubble parameter is given by
\begin{equation}\label{eq:Hubbleast}
  H_{\rm rh}=\frac{\rho_{B}^{1/2}}{\sqrt{3}\,m_P}=\sqrt{\frac{\pi^2
  g_{\ast}}{90}}\frac{T_{\rm rh}^2}{m_P}\,,
\end{equation}
where $g_{\ast}$ is the number of relativistic degrees of freedom,
which for the Standard Model in the early Universe is
$g_{\ast}=106.75$, and $T_{\rm rh}$ is the reheating temperature.
Note that $T_{\rm rh}$ is not associated with the decay of the
inflaton. It corresponds to the onset of the HBB.

It must be stressed here that the asymptotic value $\phi_F$ in the
case of a double exponential potential is not finite due to the
steepness of the potential. However, $\phi_F$ is finite if
$V(\phi)$ takes the form of the typical exponential-like uplifting
potential
\begin{equation}\label{exponentials}
  V(\phi)\simeq C_ne^{-b\phi/m_P}\,,\quad\textrm{or}\quad V(\sigma)
  \simeq C_n\sigma^{-n}\,,
\end{equation}
in terms of $\sigma$ (see Eq.~(\ref{canonfield})), where $C_n$ is
a density scale and
\begin{equation}
  b=\sqrt{2/3}\,n\,.
\end{equation}

Since its introduction in Ref.~\cite{Kachru:2003aw} to help
stabilise the volume modulus of type IIB compactifications in de
Sitter vacua, this kind of exponential-like potential has become
widespread as uplifting term. However, in our model such an
uplifting term does not stabilise the field in a minimum of the
potential. Instead, the uplifting term, when it dominates over
non-perturbative contributions at large field values, allows the
field to freeze temporarily due to excessive cosmic friction. This
excess of cosmic friction may be invoked as well when the
uplifting term cooperates with some non-perturbative contribution
to produce a minimum in the scalar potential, thus assisting
moduli stabilisation as in Ref.~\cite{Kaloper:2004yj}.

\subsection{Coincidence and constraints from BBN}
Once the field reaches its value $\phi_F$, it remains frozen until
the density of the Universe is comparable to $V(\phi_F)$. Hence,
in order for the field $\phi$ to become quintessence, we have to
constrain its evolution with the so-called \textit{coincidence
requirement}. This requires that the value at which the field
freezes, $\phi_{\rm F}$, at the beginning of the HBB is such that
\begin{equation}\label{eq:coincidence}
V(\phi_{\rm F})\simeq \Omega_{\phi}\rho_0\,,
\end{equation}
where $\Omega_{\phi}\sim 0.7$ is the ratio of the scalar density
to the critical density, and $\rho_0$ denotes the energy density
of the Universe today.

The subsequent evolution of the field depends strongly on the
value of the slope $b$ in Eq.~(\ref{exponentials})
\cite{jose,eta}. If $b^2<2$ the potential density of the scalar
field becomes dominant, and the Universe engages in a phase of
eternal acceleration. Even though string theory disfavors this
possibility due to the appearance of future horizons
\cite{Hellerman:2001yi}, this scenario cannot be ruled out from
observations. It is possible as well that the scalar density
becomes dominant without causing eternal acceleration. This occurs
for slopes in the interval $2<b^2<3(1+w_B)$, where $w_B$ is the
barotropic parameter for the background component.

If the slope of the potential lies in the interval
$3(1+w_B)<b^2<\sqrt{6}$, then the field unfreezes following an
attractor that mimics the background component. This solution has
$\rho_{\phi}/\rho_B=\textrm{const.}<1$ \cite{jose}. However, the
field does not follow this attractor immediately. It has been
shown \cite{Copeland:1997et} that the field oscillates for some
time around its attractor solution before it starts to follow its
attractor. Numerical computations \cite{Cline:2001nq} show that a
short period of accelerated expansion is possible, even in the
case of dark energy domination without eternal acceleration. In
fact, this acceleration is found when the slope $b$ in
Eq.~(\ref{exponentials}) lies in the interval
\begin{equation}\label{intacceler}
  2\lesssim b^2\lesssim 24\,,
\end{equation}
which is equivalent to $\sqrt{3}\leq n\leq 6$. More recent studies
have reduced this range somewhat. According to
Ref.~\cite{Blais:2004vt} accelerated expansion is possible only if
\mbox{$b^2<27/8$}. This would mean that brief acceleration is
possible in the range \mbox{$\sqrt 3\leq n<9/4$}. However, the
observational constraints on the density parameter and the
equation of state of Dark Energy are heavily dependent on priors
(such as a assuming a cosmological constant for the former or
primordial curvature perturbations with a constant spectral index
for the latter) as also acknowledged in Ref.~\cite{Blais:2004vt}.
This means that, were those priors modified or removed, the
allowed range of $n$ may be well enlarged. In Sec.~\ref{sec:disc}
we comment on how an exponential potential with slope $b$ in the
above interval can be theoretically motivated. Now we describe how
the coincidence requirement constrains our system once we consider
the term in Eq.~(\ref{exponentials}) as the dominant contribution
to the scalar potential.

As we said before, after the end of inflation the scalar field is
governed by Eq.~(\ref{eqnkination}) with $H=1/3t$, thus growing
with time as $\sim\ln t$. However, once the Universe becomes
radiation dominated, with $H=1/2t$, the equation of motion
features an asymptotic value for $t\to\infty$
\cite{jose,eta}. This value $\phi_F$ is given by
\begin{equation}\label{field frozen}
  \phi_F\sim\phi_{\rm e}+\sqrt{\frac{2}{3}}\,m_P\,
\textrm{ln}\left[\sqrt{\frac{90}
  {\pi^2g_{\ast}}}\frac{H_*\,m_P}{T_{\rm
  rh}^2}\right]=\phi_{\rm e}+\Delta\phi_F\,,
\end{equation}
We note here that at the end of inflation the field is very close
to the ESP, i.e. $\bar{\phi}_{\rm e}< m_P$, whereas
$\Delta\phi_{\rm F}> m_P$. Assuming also that $\phi_0\lesssim
m_P$, we approximate $\phi_F\approx \Delta\phi_{\rm F}$, with the
corresponding field $\sigma_F$ given by
\begin{equation}\label{eq:sigmafrozen}
  \sigma_F\sim \left[\sqrt{\frac{90}
  {\pi^2g_{\ast}}}\frac{H_*\,m_P}{T_{\rm
  rh}^2}\right]^{2/3}\,.
\end{equation}
Hence, given that the field has to cover a number of Planck
lengths to reach the value $\phi_F$, the non-perturbative
potential $V_{\rm np}(\phi)$ becomes subdominant with respect to
the uplift potential in Eq.~(\ref{exponentials}). Assuming that
this is the case at $\phi\sim \phi_F$, Eq.~(\ref{eq:coincidence})
applied to such a potential allows us to estimate the reheating
temperature
\begin{equation}\label{eq:reheatingtemp}
  T_{\rm rh}\sim\sqrt{H_*\,m_P}\left(\frac{\rho_0}{C_n}
  \right)^{\sqrt{3/8n^2}}\,.
\end{equation}
Requiring now the reheating temperature to be larger than the
temperature at Big Bang Nucleosynthesis (BBN), $T_{\rm BBN}$, we
obtain the bound
\begin{equation}\label{eq:upperbound}
C_n\lesssim\rho_0\left(\frac{\sqrt{H_*\,m_P}}{T_{\rm
BBN}}\right)^{2n\sqrt{2/3}}\,.
\end{equation}

\subsection{Constraints from gravitational waves}
Now we look at the constraints imposed by the density of gravitons
produced during the inflationary process. Models of quintessential
inflation are known to have a relic graviton spectrum with three
different regions. This is the result of the intermediate phase of
kination between the inflationary expansion and the HBB. Due to
the stiff equation of state during kination, the spectrum exhibits
a spike that corresponds to the production of gravitons at high
frequencies \cite{Giovannini:1999bh}.

Due to the presence of the spike at high frequencies, it is
necessary to impose an integrated bound in order
not to disturb BBN predictions. The constraint is \cite{eta}
\begin{equation}\label{eq:gravwave}
  I\equiv h^2\int_{k_{\rm BBN}}^{k_{e}}\Omega_{\rm GW}(k)\,d\ln
  k\leq2\times10^{-6}\,,
\end{equation}
where $\Omega_{\rm GW}(k)$ is the density fraction of the
gravitational waves (GWs) with \emph{physical} momentum $k$, $h=0.73$ is
the Hubble constant $H_0$ in units of $100$ km/sec/Mpc. Using the
spectrum $\Omega_{\rm GW}(k)$ as computed in
Ref.~\cite{Giovannini:1999bh}, the constrain above can be written
as \cite{eta}
\begin{equation}
  I\simeq h^2\alpha_{\rm GW}\Omega_{\gamma}(k_0)\frac{1}{\pi^3}
  \frac{H_*^2}{m_P^2}\left(\frac{H_*}{H_{\rm
  rh}}\right)^{2/3}\,,
\end{equation}
where $\alpha_{\rm GW}\simeq 0.1$ is the GW generation efficiency
during inflation, $\Omega_{\gamma}(k_0)=2.6\times 10^{-5}h^{-2}$
is the density fraction of radiation at present on horizon scales,
and ``$\textrm{rh}$'' denotes the end of kination. Making use of
Eq.~(\ref{eq:Hubbleast}) the bound above becomes
\begin{equation}
  I\simeq\frac{\alpha_{\rm GW}\Omega_{\gamma}(k_0)}{3\pi^3}
  \left(\frac{30}{\pi^2g_{\ast}}\right)^{1/3}\frac{3H_*^2}{m_P^2}
  \left(\frac{H_*m_P}{T_{\rm rh}^2}\right)^{2/3}\,.
\end{equation}
Substituting the values given above, and applying the bound in
Eq.~(\ref{eq:gravwave}) we obtain
\begin{equation}\label{eq:constr_m0}
\frac{3H_*^2}{m_P^2}
  \left(\frac{H_*m_P}{T_{\rm rh}^2}\right)^{2/3}\lesssim 4\times
  10^2\,.
\end{equation}
Using the minimum reheating temperature $T_{\rm rh}\sim T_{\rm
BBN}\sim1$ MeV we find the constraint
\begin{equation}\label{m0bound}
  H_*\lesssim 10(m_PT_{\rm BBN})^{1/2}\sim
  10^5\,\textrm{TeV}
  \,,
\end{equation}
whereas for larger reheating temperature the corresponding bound is
less demanding.

\section{Quintessence in Flux Compactifications}\label{sec:disc}
In this section we discuss whether it is possible to realise our
model of quintessential inflation with the volume modulus of type
IIB string theory. Such a realisation posseses a number of
well-known problems. One of them is related to the variation of
fundamental constants in nature. This problem can be overcome
since in the later part of the Universe history the field is
frozen, and starts evolving only today with a characteristic time
of variation within the order of the Hubble time. This fact makes
quintessence consistent with no variation in the fundamental
constants.

Another problem is that moduli fields couple to ordinary matter
with gravitational strength, therefore giving rise to long-range
forces, which are strongly constrained by observations. A possible
way out is to consider that the coupling of the
inflaton-quintessence modulus to baryonic matter is further
suppressed.\footnote{Consider, for example, a particle with mass
$m(T)$, where $T$ is the modulus in question. Expanding the mass
around the present value $T_0$ we have \mbox{$m(T)\simeq
m(T_0)+m^{_{\mbox{'}}}(T_0)\delta T$}, where the apostrophe
denotes derivative with respect to $T$. Now, one expects
\mbox{$m^{_{\mbox{'}}}(T_0)=\beta m(T_0)/m_P$}, where
\mbox{$\beta={\cal O}(1)$}. This translates into an interaction
between the particle and the modulus. Indeed, suppose that the
particle is a fermion (e.g. the electron). Then the relevant part
of the Lagrangian is \mbox{${\cal
L}_\psi=m\bar\psi\psi=m(T_0)\bar\psi\psi+ \beta[m(T_0)/m_P]\delta
T\bar\psi\psi$}. The latter term expresses an interaction between
the fermion and the modulus in question. This interaction can be
sufficiently suppressed if $\beta$ is tuned to be small:
\mbox{$\beta\lesssim 10^{-3}$}. This tuning, albeit significant,
is much less stringent than the tuning of $\Lambda$ required in
$\Lambda$CDM. We would like to thank Lorenzo Sorbo for his
feedback at this point.}

We discuss now a number of possible origins for the exponential
term $C_n\sigma^{-n}$ realising the quintessential part of the
evolution. In this paper we consider a string modulus field as
quintessence. The scalar potential of such a field is flat to all
orders in perturbation theory. Hence, a modulus field obtains its
mass solely through non-perturbative effects. Along with these
non-perturbative effects, there is a number of possible
contributions that one may consider. In type IIB
compactifications, the non-perturbative contributions to the
scalar potential $V(\phi)$ depend on the volume of the
compactified space, parametrised by the volume modulus $T$. The
realization of quintessence with this volume modulus $T$ considers
the stabilisation of the volume of the internal manifold due to
excessive cosmic friction \cite{eta,cosmofric}, rather than the
stabilisation at a minimum of the scalar potential. This, in turn,
puts forward the possibility of having a dynamical internal
manifold rather than a frozen one, as has been recently suggested
in the literature \cite{Biswas:2003kf}.

It must be stressed that the crucial event in this stabilisation
setting is the reheating of the Universe achieved through the
decay of a curvaton field. The decay products of such curvaton
field provides a thermal bath that sources the cosmic friction
which freezes the field, as explained in Sec.~\ref{sec:RaQ}.

\subsection{Through non-perturbative effects}
In type IIB compactifications, certain combinations of fluxes can
stabilise the dilaton and complex structure moduli
\cite{Giddings:2001yu}. The warp factor for antibranes can be
computed in terms of these fluxes, and it is found that this is
minimized if the antibrane is located at the tip of the
Klebanov-Strassler throat. For a set of $N$ branes sitting at the
tip of the throat the warp factor is \cite{Giddings:2001yu}
\begin{equation}
  e^{A_{\rm min}}\sim e^{-2\pi K/3Mg_s}\,,
\end{equation}
where $g_s$ is the string coupling. The $\overline{D3}$-brane
introduces an uplifting term in the scalar potential which,
written in Planck units, is given by
\begin{equation}
  \delta V(\sigma)\sim\frac{e^{4A_{\rm min}}}{\sigma^2}\equiv
  C_2\sigma^{-2}\,.
\end{equation}
This contribution dominates the scalar potential for large values
of $\sigma$ where the non-perturbative potential $V_{\rm np}$, due
to its steepness, is subdominant. Consequently, at large values of
$\sigma$ we approximate the scalar potential as
$V(\phi)\simeq\delta V$.

The reliability of the supergravity approximation in this context
requires $g_sM>1$ \cite{Klebanov:2000hb}. Also, to obtain $C_2$
small enough to comply with the coincidence requirement embodied
in Eq.~(\ref{eq:upperbound}), we must consider a choice of fluxes
such that
\begin{equation}\label{eq:quint fluxes}
K>g_s M>1\,.
\end{equation}
Tuning appropriately these values we can always generate the
appropriate scale for $C_2$. Also, in view of
Eq.~(\ref{eq:reheatingtemp}), this allows us to increase the
reheating temperature at the onset of the HBB by increasing the
units of flux K.

In this case $n=2$. Then, taking for example $H_*\sim1\,$TeV,
$T_{\rm BBN}\sim 10\,$MeV and $\rho_0\sim 10^{-120}\,m_P^4$,
Eq.~({\ref{eq:upperbound}}) gives
\begin{equation}
C_2^{1/4}\lesssim10^{-20}m_P\,.
\end{equation}
In this case, the ratio
between fluxes must be \mbox{$K/Mg_s\gtrsim22$}. Taking $g_s=0.1$, only
approximately twice as many units of $K$ flux as those of $M$ flux
are needed to generate the appropriate energy scale.

It is also possible to consider other sources for the uplifting
term $\delta V$, like for example the introduction of fluxes of
gauge field on $D7$-branes \cite{Burgess:2003ic}. In this case,
the scalar potential $V(\phi)$ is modified in a similar way,
obtaining now
\begin{equation}
\delta V\sim \frac{2\pi E^2}{\sigma^3}\equiv C_3\sigma^{-3}\,,
\end{equation}
where $E$ depends on the strength of the gauge fields considered.

In this case the constraint on $C_3$ is alleviated thanks to the
higher value of $n$. If we consider $H_*\sim1$ TeV as before, then
Eq.~({\ref{eq:upperbound}}) gives now
\begin{equation}
C_3^{1/4}\lesssim 10^{-15}m_P\,,
\end{equation}
which corresponds to an energy scale roughly within the order of
the TeV.

\subsection{In string perturbation theory}\label{sec:Qspt}
Another kind of corrections introducing an uplifting term in the
scalar potential $V(\phi)$, changing its structure at large volume
and breaking the no-scale structure, are $\alpha^{\prime}$
corrections \cite{Becker:2002nn}.

We consider the simplest case in which the volume of the
compactified space is determined by one single volume modulus $T$.
In this case, the volume of the internal manifold in the Einstein
frame is ${\cal V}=(T+\bar{T})^{3/2}$. This case has already been
considered in the literature \cite{Westphal:2005yz}. The corrected
Kh\"aler potential is
\begin{equation}\label{eq:kahlerpot}
K=K_0-2\textrm{ln}
\left(1+\frac{\hat{\xi}}{2(2\textrm{Re}\,T)^{3/2}}\right)\,,
\end{equation}
where $K_0=-3\textrm{ln}\,(T+\bar{T})$ is the tree-level Kh\"aler
potential, and $\hat{\xi}=-\frac{1}{2}\zeta(3)\chi
e^{-3\varphi/2}$ where $\chi$ is the E\"uler number of the
internal manifold, and $e^{\varphi}=g_s$ is the string coupling.
We consider the superpotential with only one non-perturbative
contribution $W=W_0+Ae^{-c\sigma}$. For large values of
$\sigma=\textrm{Re}\,T$ we can approximate the superpotential by
the classical one $W=W_0$. In this case, the scalar potential
computed from $W=W_0$ using the K\"ahler potential in
Eq.~(\ref{eq:kahlerpot}) is
\begin{equation}
V\approx\delta
V\sim\frac{\hat{\xi}\,W_0^2}{(\textrm{Re}\,T)^{9/2}}\,,
\end{equation}
for the tree level potential is exponentially suppressed for large
values of $\sigma$ [c.f. Eq.~(\ref{scalarpot0})]. When
$(\textrm{Re}\,T)^{3/2}$ is large compared to $\hat{\xi}$ we can
compute the canonically normalized field $\phi$ using $K\approx
K_0$. As a result, $\phi$ is still given by
Eq.~(\ref{canonfield}), and consequently we can identify the
exponential potential in terms of $\sigma$
\begin{equation}
  \delta V\sim\frac{\hat{\xi}\,W_0^2}{\sigma^{9/2}}\equiv
  C_{9/2}\sigma^{-9/2}\,.
\end{equation}
In this case, the bound for $C_{9/2}$ obtained using
Eq.~(\ref{eq:upperbound}) with $H_*\sim1$ TeV results in
\begin{equation}
  (C_{9/2})^{1/4}\lesssim 10^{-7}m_P\,,
\end{equation}
where the upper bound corresponds to the intermediate scale
$C_{9/2}^{1/4}\sim10^{11}$ GeV.

\subsection{Constraints on the volume modulus}
Not only must this picture be consistent from the theoretical
point of view, but also from the observational one. In this sense,
it is well known that the excessive production of light moduli
fields with a late decay can spoil the abundance of light elements
predicted by BBN. Here, we are interested in the risks put forward
by the volume modulus. The most evident one is that, given that
this field controls the volume of the compactified space,
Kaluza-Klein modes may jeopardize BBN predictions.

In closed string theory, in addition to the usual translational
modes, vibrational modes of closed strings wrapping around the
compact manifold are also present. The mass of these excitations
follows an inverse relation to the radius of the compactification
\cite{Conlon:2005ki}
\begin{equation}\label{eq:KKmodes}
  m_{\rm KK}\sim \frac{m_s}{R}\sim \frac{g_s \,m_P}{{\cal V}^{2/3}_{\rm st}}\,,
\end{equation}
where $R$ is the radius of the compactified space defined by the
volume of the Calabi-Yau manifold ${\cal V}_{\rm st}\sim R^{6}$,
$g_s$ is the string coupling and $m_s$ the string mass scale.

We now constrain the volume of the compactified space in order to
avoid a late decay disturbing the abundances of light elements
predicted by BBN. Recalling now Eq.~(\ref{eq:sigmafrozen}) for the
frozen field, we can estimate the volume ${\cal V}_{\rm
st}(\sigma_F)$, with which we can compute the mass of the lightest
KK modes, as given by Eq.~(\ref{eq:KKmodes}). Also, assuming
interaction of at least gravitational strength, these modes decay
at the temperature
\begin{equation}
  T_{\rm KK}\sim\sqrt{\Gamma\, m_P}\sim\frac{g_s^{3/2}m_P}{\sigma_F^{3/2}}\,,
\end{equation}
where we have considered the decay rate $\Gamma\sim m_{\rm
KK}^3/m_P^2$. In order to preserve the abundances predicted by BBN
we require $T_{\rm KK}>T_{\rm BBN}$. The frozen field value
$\sigma_F$ is thus forced to satisfy
\begin{equation}
  \sigma_F<\left(\frac{g_s^{3/2}m_P}{T_{\rm BBN}}\right)^{2/3}\,.
\end{equation}
Applying this bound to the expression in
Eq.~(\ref{eq:sigmafrozen}) for the frozen field results in a bound
for the reheating temperature $T_{\rm rh}$
\begin{equation}
  T_{\rm rh}>\left[\frac{30}{\pi^2g_{\ast}g_s^3}\right]^{1/4}
\sqrt{T_{\rm BBN}\,H_*}
  \sim\sqrt{T_{\rm BBN}\,H_*}\,,
\end{equation}
which for $H_*\sim\,1\,\textrm{TeV}$ and $T_{\rm BBN}\sim\,1$ MeV
results in $T_{\rm rh}>1$ GeV. Taking then the reheating
temperature $T_{\rm rh}\sim\,10$ GeV alleviates the constraint on
$H_*$ imposed by Eq.~(\ref{eq:constr_m0}), which now becomes
\begin{equation}
  H_*\lesssim 10^7\,\textrm{TeV}\,.
\end{equation}
However, the reheating temperature results in a significantly
stronger bound on the coefficient $C_{9/2}$. Using
Eq.~(\ref{eq:reheatingtemp}) with $T_{\rm rh}\sim10$ GeV we obtain
\begin{equation}
  C_{9/2}^{1/4}\lesssim10^{-12}m_P\sim10^6\textrm{ GeV}\,.
\end{equation}

\section{Summary and Conclusions}\label{sec:SaC}

We have investigated in detail a realisation of quintessential
inflation within the framework of string theory. Our inflaton -
quintessence field is a string modulus with a characteristic
runaway scalar potential. In our treatment we have avoided to
specify in full detail our string inspired theoretical basis in
order for our considerations to remain as generic as possible. In
that sense, our quintessential inflation realisation is more like
a paradigm than a specific model.

In this spirit we considered a broadly accepted form of the
non-perturbative superpotential (which can be due to, for example,
gaugino condensation or $D$-brane instantons) and of the
tree-level K\"{a}hler potential shown in Eqs.~(\ref{W}) and
(\ref{eq:treeKahler}) respectively. The resulting non-perturbative
potential appears in Eq.~(\ref{scalarpot0}) and is of the form
\mbox{$V_{\rm np}(\sigma)\propto \sigma^{-\ell}e^{-\mu c\sigma}$},
where \mbox{$\sigma=$ Re $T$} and $\ell,\mu=1,2$ depending which
term dominates. Due to Eq.~(\ref{eq:treeKahler}) the canonically
normalised modulus $\phi$ is in fact related to $\sigma$ as
\mbox{$\sigma(\phi)=e^{\sqrt{\frac{2}{3}}\phi/m_P}$} [c.f.
Eq.~(\ref{canonfield})]. Hence, the potential $V_{\rm np}(\phi)$
is, in fact, of the form of a double exponential.

Such a potential is too steep to support inflation. This is why,
we have assumed the existence of an enhanced symmetry point (ESP)
at some value $\phi_0$, corresponding to potential density
\mbox{$V_0=3H_*^2m_P^2$}. The most salient effect of the ESP is
that it can trap the rolling modulus and effectively stop it from
rolling. Indeed, as the modulus crosses the ESP, particle
production can generate a contribution $V_{\rm int}(\phi)$ to the
scalar potential able to stop the roll of the modulus and trap the
latter at the ESP. However, apart from the phenomenon of particle
production described in \cite{Kofman:2004yc}, in this paper we
have also taken into account that the ESP generates a `flat patch'
on the scalar potential because of the condition
\mbox{$V'(\phi_0)=0$}. Hence, all things considered, the trapping
mechanism that operates at the ESP may set the field at a locally
flat region in field space. In this paper we show how, after the
field is released from its trapping, this flatness may result in
enough inflation to solve the flatness and horizon problems.

After being trapped, the modulus oscillates around the ESP under
the influence of $V_{\rm int}(\phi)$. These oscillations deplete
its initial kinetic density until the latter decreases below
$V_0$, in which case a phase of so-called trapped inflation begins
\cite{Kofman:2004yc}. Trapped inflation dilutes $V_{\rm int}$
until the latter is unable to restrain the field, at which time
the modulus is released and continues rolling over $V(\phi)$. As
we have shown, trapped inflation is brief and cannot suffice for
the solution of the horizon and flatness problems. The duration of
trapped inflation is independent from the value of the initial
kinetic density of the modulus, because all excess of the latter
is depleted before the onset of trapped inflation. Hence, our
framework is largely independent of initial conditions, provided
trapping occurs.

To study the dynamics after the field is released from its trapping,
we have followed a phenomenological approach. In it, the scalar
potential $V(\sigma)$ consists of a `background' potential $V_{\rm
np}(\sigma)$ of non-perturbative origin (from gaugino condensation,
for example), and a phenomenological potential $V_{\rm ph}(\sigma)$
chosen so that the scalar potential $V(\sigma)=V_{\rm
np}(\sigma)+V_{\rm ph}(\sigma)$ satisfies $V^{\prime}(\phi_0)=0$.
The guiding line to pick out $V_{\rm ph}$ is to demand that it
decreases faster than $V_{\rm np}$, so that the former is only
important when the field is close to the ESP. Given the
characteristic form of $V_{\rm np}(\sigma)$, the simplest choice for
$V_{\rm ph}$ is to take $V_{\rm ph}\propto e^{-q\sigma}$ far from
the ESP with $q>\mu c$ a constant. A particular realisation of
$V_{\rm ph}$ is given in Eq.~(\ref{eq:vph}). In this setting, we
show in Appendix \ref{app:minapp} that the terms of order forth and
higher in a series expansion around the ESP become important only
after the end of inflation. Therefore, the inflationary dynamics can
be fully accounted for through the scalar potential
$V(\phi)=V_0+\frac{1}{2}m_{\phi}^2\bar{\phi}^2+\frac{1}{3!}\vppp
\bar{\phi}^3$, where $\bar{\phi}\equiv\phi-\phi_0$.

In this paper we are concerned with the observable amount of
inflation. We thus pay special attention to the last $N\approx60$
$e$-foldings of inflation, and compute the curvature perturbation,
$\pphi$, generated by the inflaton field when the observable
Universe leaves the horizon. We show how in our model $\vppp$ is
fully determined by $H_*$ and ${\cal P}^{1/2}_{\phi}$,
Eq.~(\ref{eq:pert}). This offers two significant advantages. First,
the dynamics becomes described in terms of simple and meaningful
physical quantities relevant to inflation. Second, the parameter
space of the model becomes automatically constrained by the observed
value of the curvature perturbation, which $\pphi$ cannot exceed,
i.e. $\pphi\lesssim4.8\times10^{-5}$. Note that the observed
curvature perturbation in our model is generated by a curvaton field
which is also employed to reheat the Universe.

A side effect of our model is that it imposes a lower bound in
$\pphi$ [c.f. Eq.~(\ref{eq:intpert})], which in turn translates into
an upper bound on $H_*$, Eq.~(\ref{eq:upperm0}). We recognise these
bounds as an artifact of the model, that appears when we parametrise
our ignorance about the behaviour of $V(\phi)$ around the ESP. A
simple way around these unphysical bounds is to let $q$ have a
running.

We find that we can attain enough inflation to solve the flatness
and horizon problems if $|m_{\phi}|\lesssim H_*$ and $\vppp\ll H_*$.
In particular, we discover that the field may have two different
inflationary histories, or regimes, ending in a phase of slow-roll
during which the observable Universe leaves the horizon. Whereas one
of the inflationary regimes consists only of slow-roll inflation,
the other regime involves a previous phase of eternal inflation.
Eqs.~(\ref{eq:intsecsr2}) and (\ref{eq:inteternal}) [c.f.
Eq.~(\ref{eq:f(g)})] determine which of these inflationary regimes
will be driven by the field. Of course, consistency requires that
the regimes must be compatible with the trapping condition,
Eq.~(\ref{eq:trp}). The values of $g$ corresponding to the different
inflationary regimes, and to the trapping condition, are pictured in
Figs.~\ref{fig:infmap1} and \ref{fig:infmap2}, corresponding to
$m_{\phi}^2<0$ and $m_{\phi}^2\geq0$ respectively. The results in
both cases are qualitatively the same: the inflationary regime
involving eternal inflation is always compatible with trapping.
However, the inflationary regime where only slow-roll inflation
occurs, is compatible with trapping only if $|\eta|$ when
sufficiently large, Eq.~(\ref{eq:eta}). Finally, an acceptable
inflationary cosmology is also achievable through fast-roll
inflation. This possibility arises only when the ESP results in a
maximum with $|m_{\phi}^2|\sim H_*^2$. In this case, $H_*$ must
satisfy an upper bound, Eq.~(\ref{eq:Hfr}), so that the field does
not generate an excessive amount of perturbations.

After inflation is terminated the field becomes dominated by its
kinetic density and the Universe enters a period of so-called
kination \cite{kination}. During this period the modulus is
oblivious of the scalar potential and its rapid roll can send it
very far in field space. To reheat the Universe we have assumed a
curvaton field, whose decay products in time dominate the kinetic
density of the rolling modulus \cite{curvreh}. The existence of a
suitable curvaton, apart from reheating the Universe, can also
provide the required density perturbations \cite{curv} and allow for
a low inflation scale \cite{orth,low}. After reheating, the modulus
roll is inhibited by cosmological friction as discussed in
Refs.~\cite{eta,cosmofric}. Consequently, in little time, the
modulus freezes at a value $\sigma_F$ shown in
Eq.~(\ref{eq:sigmafrozen}). The modulus remains frozen until the
present, which guarantees the non-variation of the fundamental
constants through-out the hot big bang despite the fact that the
modulus is not stabilised in a local minimum.
Eq.~(\ref{eq:sigmafrozen}) shows that $\sigma_F$ depends on the
reheating temperature, which, in turn, depends on the details of the
curvaton model. Therefore, in our framework curvaton physics can
determine the value of the string modulus in our Universe.

After the end of kination the modulus is expected to freeze at a
large value at the tail of the scalar potential. At such values, we
assume that the latter may be dominated by a contribution of the
form \mbox{$V(\sigma)=C_n/\sigma^n$} [c.f.
Eq.~(\ref{exponentials})], where $C_n$ is a density scale and the
value of $n$ is model dependent. The value of $C_n$ is determined by
the coincidence requirement, if the modulus is to become
quintessence at present. This requirement results in the constraint
in Eq.~(\ref{eq:reheatingtemp}), where we see that $C_n$ depends
also on the reheating temperature and, therefore, it is determined
by curvaton physics. The requirement that reheating occurs before
Big Bang Nucleosynthesis (BBN) sets an upper bound on $C_n$, shown
in Eq.~(\ref{eq:upperbound}). Crucially, the bound is strongly
$n$-dependent.

This type of uplifting potential is again broadly accepted in
string flux compactifications.We have discussed some possibilities
for the origin of the uplifting potential, such as RR and NS
fluxes on $\overline{D3}$-branes (where $n=2$ and
$C_2^{1/4}\lesssim 10$~MeV), gauge field fluxes on $D7$-branes
(where $n=3$ and $C_3^{1/4}\lesssim 1$~TeV) and
$\alpha'$-corrections (where $n=9/2$ and $C_{9/2}^{1/4}\lesssim
10^{11}$~GeV). $C_n$ may be further constrained in order to
protect BBN from heavy KK modes. It turns out that this constraint
is important only in the $n=9/2$ case where the bound on $C_{9/2}$
is strengthened to $C_{9/2}^{1/4}\lesssim 10^6$~GeV.\footnote{The
uplifting potential considered may even have a perturbative
origin, due to duality transformations $\sigma\rightarrow
1/\sigma$ of polynomial terms at large values. KD wishes to thank
P.M.~Petropoulos for pointing this out.} Note that the above
bounds are much more reasonable than the fine-tuning of $\Lambda$
required in $\Lambda$CDM.

According to Eq.~(\ref{exponentials}), the scalar potential for the
canonically normalised field $\phi$ is of exponential form. For the
above discussed values of $n$ and according to
Ref.~\cite{Cline:2001nq} the modulus can cause a brief period of
accelerating expansion while it unfreezes at present \cite{jose}.
Thus, our model does not lead to eternal acceleration and therefore
does not suffer from the existence of future horizons
\cite{Hellerman:2001yi}. In the future, the modulus eventually
approaches $+\infty$ which corresponds to a supersymmetric ground
state. If $\sigma$ is the volume modulus then this final state leads
to decompactification of the extra dimensions.

Some numerical simulations more recent than Ref.~\cite{Cline:2001nq}
suggest that the parameter space for brief acceleration is somewhat
reduced. In particular, in Ref.~\cite{LopesFranca:2002ek} the
authors demonstrate that brief acceleration can indeed take place in
the interval \mbox{$\sqrt 2<b\leq\sqrt 3$} corresponding to the
range \mbox{$\sqrt 3<n\leq 3/\sqrt 2$}, which includes the case
$n=2$. This is confirmed by Refs.~\cite{Kallosh:2002gf} and
\cite{Kehagias:2004bd}, where it is implied that brief acceleration,
which is terminated at present, can be attained even for larger
values of $b$. In Ref.~\cite{Blais:2004vt} it is indeed shown that
one may have brief acceleration up to the value
\mbox{$b=\frac{3}{4}\sqrt 6$} corresponding to \mbox{$n=9/4$}. The
authors of Ref.~\cite{Blais:2004vt}  acknowledge the fact that the
parameter space determined by observations is affected by assumed
priors and may be extended further. Hence, the case \mbox{$n=3$}
might still be successful. However, in view of the above, the case
of $\alpha'$-corrections to the K\"{a}hler potential
(Sec.~\ref{sec:Qspt}) is probably excluded.

In conclusion, we have investigated a paradigm of quintessential
inflation using a sting modulus as the inflaton - quintessence
field. We have shown that there exists ample parameter space for
successful quintessential inflation to occur with reasonable
values of the model parameters (with possibly a mild tuning of
gravitational couplings between the modulus and the standard model
fields, required to overcome the 5th force problem). The
possibility of attaining an acceptable inflationary cosmology
depends crucially on the values of $\vppp$ and $m_{\phi}$ relative
to the Hubble scale $H_*$ during inflation. The latter is
determined by the location of the ESP and can be much different to
$m_\phi$ or $\vppp$. Furthermore, a crucial role is played by the
value of the reheating temperature, which controls both the value
of the frozen modulus and the required value of the density scale
in the uplift potential. The reheating temperature is determined,
in turn, by the particulars of the assumed curvaton field. Because
the modulus remains frozen until the present, the model
may conceivably be linked to the intriguing possibility that some
of the fundamental constants, such as the fine-structure constant,
may begin to vary at present.

\section{Acknowledgments} We wish to thank James Gray for useful
comments and Lorenzo Sorbo for stimulating discussions. KD wishes to
thank Steve~Giddings and Marios~Petropoulos for interesting
discussions and the Aspen Center for Physics for the hospitality.
This work was supported (in part) by the European Union through the
Marie Curie Research and Training Network "UniverseNet"
(MRTN-CT-2006-035863) and by PPARC (PP/D000394/1). KD is supported
by PPARC grant PPA/Y/S/2002/00272 and EU grant MRTN-CT-2004-503369.

\section{Appendix}\label{sec:app}
\subsection{Elucidating the oscillatory regime}\label{trapdynamics}
The key point to solve the oscillatory regime is to determine how
the mass term $g^2\phi^2$ decreases in relation to the kinetic term
$k^2$ (see Eq.~(\ref{eq:fluct})) once particle production finishes.

After the field crosses the ESP for the first time, the occupation
number becomes $n_k^{(1)}=\exp(-\pi k^2/g\dot{\phi}_0)$. During
particle production we assume the same growth of the occupation
number in all of the excited $\chi$ modes every time the field
crosses the ESP. In this case, the occupation number at the end of
particle production becomes $n_k^{\rm end}\sim g^{-2}\exp(-\pi
k^2/g\dot{\phi}_0)$.

In this process of particle production the expansion of the Universe
is neglected. Therefore, if we consider the expansion of the
Universe after particle production, the occupation number for a
$\chi$-mode with \emph{physical} momentum $k$ becomes
\begin{equation}
  n_k^{\rm end}\sim g^{-2}\exp(-\pi (ka)^2/g\dot{\phi}_0)\,.
\end{equation}

Let us now consider $a\simeq 1$. Given that $m^2\gg H^2$ at the end
of particle production, we can approximate the integral
\begin{equation}
  \langle\chi^2\rangle\propto \int_0^{\infty}\frac{n_k^{\rm
  end}k^2dk}{\sqrt{k^2+g^2\phi^2}}
\end{equation}
by replacing the field $\phi$ by its average
$\overline{\phi}(t)\approx\Phi(t)$. Now, the main contribution to
the average expectation value $\langle\overline{\chi^2}\rangle$ for
$a\simeq 1$, comes from momenta $k^2\sim(g\dot{\phi}_0)\sim
g^2\Phi^2$. Therefore, we can approximate the integral by absorbing
the term $g^2\Phi^2$ into $k^2$, namely $k^2+g^2\Phi^2\approx
2k^2\sim k^2$, thus obtaining
\begin{equation}\label{eq:comfluct}
  \langle\overline{\chi^2}(t)\rangle\sim g^{-2} \int_0^{\infty}
\textrm{exp}(-\pi(ka)^2/g\dot{\phi}_0)\,k\,dk\propto a^{-2}\,.
\end{equation}

On the other hand, given that the $\chi$ particles become
relativistic at the end of particle production, their average
density $\overline{\rho}_{\chi}\sim n_{\chi}\sqrt{k^2+g^2\Phi^2}$
must scale as $a^{-4}$. We thus conclude that $\Phi$ must initially
scale at least as fast as $a^{-1}$, for otherwise the mass term
$g^2\Phi^2$ would dominate over $k^2$ in the average density
$\overline{\rho}_{\chi}$.

To compute the initial scaling law for $\Phi$ we compute the
depletion rate of the modulus density in the time-dependent
quadratic potential $V_{\rm
int}(\phi,\chi)=\frac{1}{2}g^2\langle\chi^2\rangle\phi^2$. Owing to
$m^2> H^2$, the depletion rate of the energy of the oscillations
$\rho_{\rm osc}$ can be computed following
Ref.~\cite{Turner:1983he}. We find
\begin{equation}
  \rho_{\rm
osc}\propto a^{-4}\,.
\end{equation}
Writing now $\rho_{\rm osc}=V_{\rm int}(\Phi)\propto a^{-2}\Phi^2$
we conclude that
\begin{equation}
  \Phi\propto a^{-1}\,.
\end{equation}

This result is valid just for $a\simeq 1$. However, this scaling for
$\Phi$ implies that $\langle\chi^2\rangle$ remains independent of
$\phi$, which in turn prevents $\Phi$ from decreasing faster than
$a^{-1}$. If $\Phi$ decreased slower than $a^{-1}$, then the term
$g^2\Phi^2$ would come to dominate $k^2$. In this case we have
$\langle\chi^2\rangle\propto n_{\chi}\phi^{-1}$, and again a
time-dependent linear potential in $\phi$: $V_{\rm int}\propto
n_{\chi}\phi$. Following Ref.~\cite{Turner:1983he}, we obtain that
the amplitude $\Phi$ in this case scales exactly as before, i.e.
\mbox{$\Phi\propto a^{-1}$}.

We conclude that as soon as particle production finishes, the
amplitude of oscillations $\Phi$ and the expectation value
$\langle\overline{\chi^2}\rangle$ scale as follows
\begin{equation}
  \Phi\propto
  a^{-1}\quad\textrm{and}\quad\langle\overline{\chi^2}\rangle\propto
  a^{-2}\,,
\end{equation}
and the regime is maintained as long as $m^2> H^2$.

\subsection{Minimal realisation of an ESP}\label{app:minapp}
Our minimal realisation is based on the condition that the
phenomenological term $V_{\rm ph}$ becomes subdominant with
respect to the non-perturbative contribution $V_{\rm
np}\propto\sigma^{-\ell}e^{-\mu c\sigma}$ when the field is away
from the symmetry point. This means that $V_{\rm ph}$ must at
least decrease exponentially when it starts to become subdominant.
A particular choice to fulfill this is
\begin{equation}\label{eq:vph}
V_{\rm ph}(\sigma)=-\frac{M_{\rm
ph}}{\cosh^{-q}(\sigma-\Sigma)}\,,
\end{equation}
which results in the scalar potential
\begin{equation}
V(\sigma)\approx M_{\rm np}\sigma^{-\ell}e^{-\mu c\sigma}-M_{\rm
ph}\cosh^{-q}(\sigma-\Sigma)\,.
\end{equation}
To account now for an ESP we impose
\begin{equation}\label{eq:esp1}
\frac{\partial
V}{\partial\sigma}\Big|_{\sigma_0}=0\quad\textrm{and}\quad
\frac{\partial^2V}{\partial\sigma^2}\Big|_{\sigma_0}=
\frac{3}{2}\frac{m_{\phi}^2}{\sigma_0^2}\,,
\end{equation}
where $m_{\phi}^2=V^{\prime\prime}_0$. To have a positive density
at the ESP while satisfying Eqs.~(\ref{eq:esp1}), we must impose
$\sqrt{q}>\mu c$ (typical values are $c\lesssim1$, so one may take
$q\gtrsim1$). We then describe the parameter space of the model
through the variable
\begin{equation}\label{eq:variable}
  \xi\equiv\frac{\sqrt{q}}{c\mu}>1\,.
\end{equation}
This quantity determines how close to the ESP the potential
$V_{\rm ph}(\sigma)$ becomes exponentially suppressed with respect
to $V_{\rm np}(\sigma)$. The scaling of these two contributions
with $\sigma$ may be estimated as
\begin{equation}
V_{\rm np}(\sigma)\propto e^{-\mu
c\bar\sigma}\left(\frac{\sigma_0}{\sigma}\right)^{\ell}
\quad,\quad V_{\rm ph}(\sigma)\propto e^{-q\bar\sigma}\,.
\end{equation}
where $\bar\sigma\equiv\sigma-\sigma_0$. Therefore, writing
$r\equiv V_{\rm ph}(\sigma)/V_{\rm np}(\sigma)$, we have
\begin{equation}
\frac{r(\sigma)}{r(\sigma_0)}\sim\left(1+\frac{\bar\sigma}{\sigma_0}\right)^{\ell}
\displaystyle e^{-(q-\mu c)\bar\sigma}\,.
\end{equation}
The phenomenological term becomes negligible for
$\bar{\sigma}>\bar{\sigma}_{\rm ph}$, where \mbox{$\bar{\sigma}_{\rm
ph}\equiv(q-\mu c)^{-1}\gtrsim q^{-1}$}. Switching to the
canonically normalised field $\phi$ we obtain
\begin{equation}\label{phiph0}
\bar{\phi}_{\rm ph}=
\sqrt{3/2}\ln\left(1+\frac{1}{q\sigma_0}\right)m_P
\sim\frac{m_P}{\xi^2(c\mu)^2\sigma_0}\,,
\end{equation}
which, qualitatively, is the expected result; the larger the $\xi$
is the closer to the ESP the $V_{\rm ph}$ becomes suppressed with
respect to $V_{\rm np}$.

Now, in terms of this $\xi$ we obtain
\begin{equation}\label{eq:third2}
   V^{\textrm{\tiny(3)}}_0\approx-2(\sqrt{2/3}
   \mu)^3(c\sigma_0)^3\xi^2\frac{3H_*^2}{m_P}+\sqrt{6}
   \frac{m_{\phi}^2}{m_P}\,.
\end{equation}
Taking $\xi^2>1$, and using the slow-roll condition $|\eta|<1$, the
second term in the right-hand side of the above expression becomes
negligible. Hence,
\begin{equation}\label{eq:thirdderiv3}
  |V^{\textrm{\tiny{(3)}}}_0|>(c\sigma_0)^3
  \frac{3H_*^2}{m_P}\,,
\end{equation}

With this particular realisation we may compute when terms of
higher order must be considered. Terms of higher order start to
become important when $|\vppp|\sim|\vpppp|\bar{\phi}_{\rm HO}$.
Computing $\vpppp$ this occurs for $\bar{\phi}_{\rm
HO}\sim\xi^{-1}m_P/(c\sigma_0)$. At the end of inflation
$|\eta|\sim1$, and the field value is of order $\bar{\phi}_{\rm
e}\sim H_*^2/|\vppp|$. Using the expression above we have
\[
\bar{\phi}_{\rm HO}\sim\xi(c\sigma_0)^2\bar{\phi}_{\rm
e}>\bar{\phi}_{\rm e}\,.
\]
Thus, within our minimal approach the inflationary dynamics may be
accounted for using the scalar potential $V(\phi)$ in
Eq.~(\ref{eq:scpot}).

\begin{thebiblio}{03}

{\small

\bibitem{all}
S.~Perlmutter {\it et al.}  
Astrophys.\ J.\  {\bf 517} (1999) 565;
M.~Tegmark {\it et al.}  
Phys.\ Rev.\ D {\bf 69} (2004) 103501;
M.~Colless,
astro-ph/0305051;
 D.~N.~Spergel {\it et al.},
astro-ph/0603449;
W.~L.~Freedman, J.~R.~Mould, R.~C.~.~Kennicutt and B.~F.~Madore,
astro-ph/9801080.

\bibitem{L}
T.~Padmanabhan,
Phys.\ Rept.\  {\bf 380} (2003) 235;
S.~Weinberg,
Rev.\ Mod.\ Phys.\  {\bf 61} (1989) 1.

\bibitem{ed}
E.~J.~Copeland, M.~Sami and S.~Tsujikawa,
hep-th/0603057.

\bibitem{early}
P.~J.~E.~Peebles and B.~Ratra,
Astrophys.\ J.\  {\bf 325} (1988) L17;
B.~Ratra and P.~J.~E.~Peebles,
Phys.\ Rev.\ D {\bf 37} (1988) 3406.

\bibitem{Q}
L.~M.~Wang, R.~R.~Caldwell, J.~P.~Ostriker and P.~J.~Steinhardt,
Astrophys.\ J.\  {\bf 530} (2000) 17;
I.~Zlatev, L.~M.~Wang and P.~J.~Steinhardt,
Phys.\ Rev.\ Lett.\  {\bf 82} (1999) 896;
G.~Huey, L.~M.~Wang, R.~Dave, R.~R.~Caldwell and P.~J.~Steinhardt,
Phys.\ Rev.\ D {\bf 59} (1999) 063005;
R.~Caldwell, R.~Dave and P.~J.~Steinhardt,
Phys.\ Rev.\ Lett.\  {\bf 80} (1998) 1582.

\bibitem{Guth:1980zm}
A.~H.~Guth,
Phys.\ Rev.\ D {\bf 23} (1981) 347.

\bibitem{quinf}
P.~J.~E.~Peebles and A.~Vilenkin,
Phys.\ Rev.\ D {\bf 59} (1999) 063505.

\bibitem{QI}
R.~A.~Frewin and J.~E.~Lidsey,
Int.\ J.\ Mod.\ Phys.\ D {\bf 2} (1993) 323;
S.~C.~C.~Ng, N.~J.~Nunes and F.~Rosati,
Phys.\ Rev.\ D {\bf 64} (2001) 083510;
W.~H.~Kinney and A.~Riotto,
Astropart.\ Phys.\  {\bf 10} (1999) 387;
M.~Peloso and F.~Rosati,
JHEP {\bf 9912} (1999) 026;
K.~Dimopoulos,
Nucl.\ Phys.\ Proc.\ Suppl.\  {\bf 95} (2001) 70;
G.~Huey and J.~E.~Lidsey,
Phys.\ Lett.\ B {\bf 514} (2001) 217;
A.~S.~Majumdar,
Phys.\ Rev.\ D {\bf 64} (2001) 083503;
N.~J.~Nunes and E.~J.~Copeland,
Phys.\ Rev.\ D {\bf 66} (2002) 043524;
G.~J.~Mathews, K.~Ichiki, T.~Kajino, M.~Orito and M.~Yahiro,
astro-ph/0202144;
M.~Sami and V.~Sahni,
Phys.\ Rev.\ D {\bf 70} (2004) 083513;
A.~Gonzalez, T.~Matos and I.~Quiros,
Phys.\ Rev.\ D {\bf 71} (2005) 084029;
G.~Barenboim and J.~D.~Lykken,
Phys.\ Lett.\ B {\bf 633} (2006) 453;
V.~H.~Cardenas,
Phys.\ Rev.\ D {\bf 73} (2006) 103512;
B.~Gumjudpai, T.~Naskar and J.~Ward,
JCAP {\bf 0611} (2006) 006.

\bibitem{jose}
K.~Dimopoulos and J.~W.~F.~Valle,
Astropart.\ Phys.\  {\bf 18} (2002) 287.

\bibitem{eta}
K.~Dimopoulos,
Phys.\ Rev.\ D {\bf 68} (2003) 123506.

\bibitem{BuenoSanchez:2006eq}
  J.~C.~Bueno Sanchez and K.~Dimopoulos,
  Phys.\ Lett.\ B {\bf 642} (2006) 294.

\bibitem{Kofman:2004yc}
  L.~Kofman, A.~Linde, X.~Liu, A.~Maloney, L.~McAllister and E.~Silverstein,
  JHEP {\bf 0405} (2004) 030.

\bibitem{kination}
B.~Spokoiny,
Phys.\ Lett.\ B {\bf 315} (1993) 40;
M.~Joyce and T.~Prokopec,
Phys.\ Rev.\ D {\bf 57} (1998) 6022.

\bibitem{cosmofric}
N.~Kaloper and K.~A.~Olive,
Astropart.\ Phys.\  {\bf 1} (1993) 185.
R.~Brustein, S.~P.~de Alwis and P.~Martens,
Phys.\ Rev.\ D {\bf 70} (2004) 126012;
T.~Barreiro, B.~de Carlos, E.~Copeland and N.~J.~Nunes,
Phys.\ Rev.\ D {\bf 72} (2005) 106004.

\bibitem{curv}
D.~H.~Lyth and D.~Wands,
Phys.\ Lett.\ B {\bf 524} (2002) 5.
%
K.~Enqvist and M.~S.~Sloth,
Nucl.\ Phys.\ B {\bf 626} (2002) 395;
T.~Moroi and T.~Takahashi,
Phys.\ Lett.\ B {\bf 522} (2001) 215
[Erratum-ibid.\ B {\bf 539} (2002) 303];
S.~Mollerach,
Phys.\ Rev.\ D {\bf 42} (1990) 313.

\bibitem{curvreh}
B.~Feng and M.~z.~Li,
Phys.\ Lett.\ B {\bf 564} (2003) 169;
A.~R.~Liddle and L.~A.~Urena-Lopez,
Phys.\ Rev.\ D {\bf 68} (2003) 043517;
C.~Campuzano, S.~del Campo and R.~Herrera,
Phys.\ Lett.\ B {\bf 633} (2006) 149.

\bibitem{sneu}
J.~McDonald,
Phys.\ Rev.\ D {\bf 68} (2003) 043505;
A.~Mazumdar and A.~Perez-Lorenzana,
Phys.\ Rev.\ Lett.\  {\bf 92} (2004) 251301;
J.~McDonald,
Phys.\ Rev.\ D {\bf 70} (2004) 063520.

\bibitem{mssm}
K.~Enqvist and A.~Mazumdar,
Phys.\ Rept.\  {\bf 380} (2003) 99;
K.~Enqvist, S.~Kasuya and A.~Mazumdar,
Phys.\ Rev.\ Lett.\  {\bf 90} (2003) 091302;
K.~Enqvist, A.~Jokinen, S.~Kasuya and A.~Mazumdar,
Phys.\ Rev.\ D {\bf 68} (2003) 103507;
S.~Kasuya, M.~Kawasaki and F.~Takahashi,
Phys.\ Lett.\ B {\bf 578} (2004) 259;
K.~Enqvist, S.~Kasuya and A.~Mazumdar,
Phys.\ Rev.\ Lett.\  {\bf 93} (2004) 061301;
K.~Enqvist,
Mod.\ Phys.\ Lett.\ A {\bf 19} (2004) 1421;
R.~Allahverdi, K.~Enqvist, A.~Jokinen and A.~Mazumdar,
JCAP {\bf 0610} (2006) 007.

\bibitem{nmssm}
M.~Bastero-Gil, V.~Di Clemente and S.~F.~King,
Phys.\ Rev.\ D {\bf 67} (2003) 103516;
Phys.\ Rev.\ D {\bf 67} (2003) 083504;
M.~Postma,
Phys.\ Rev.\ D {\bf 67} (2003) 063518;
K.~Hamaguchi, M.~Kawasaki, T.~Moroi and F.~Takahashi,
Phys.\ Rev.\ D {\bf 69} (2004) 063504;
J.~McDonald,
Phys.\ Rev.\ D {\bf 69} (2004) 103511.

\bibitem{pngb}
K.~Dimopoulos, D.~H.~Lyth, A.~Notari and A.~Riotto,
JHEP {\bf 0307} (2003) 053;
E.~J.~Chun, K.~Dimopoulos and D.~Lyth,
Phys.\ Rev.\ D {\bf 70} (2004) 103510;
R.~Hofmann,
Nucl.\ Phys.\ B {\bf 740} (2006) 195.

\bibitem{orth}
K.~Dimopoulos and G.~Lazarides,
Phys.\ Rev.\ D {\bf 73} (2006) 023525.

\bibitem{PQ}
K.~Dimopoulos, G.~Lazarides, D.~Lyth and R.~Ruiz de Austri,
JHEP {\bf 0305} (2003) 057.

\bibitem{liber}
K.~Dimopoulos and D.~H.~Lyth,
Phys.\ Rev.\ D {\bf 69} (2004) 123509.

\bibitem{low}
K.~Dimopoulos, D.~H.~Lyth and Y.~Rodriguez,
JHEP {\bf 0502} (2005) 055;
K.~Dimopoulos,
Phys.\ Lett.\ B {\bf 634} (2006) 331.

\bibitem{Hull:1995mz}
  C.~M.~Hull and P.~K.~Townsend,
  Nucl.\ Phys.\ B {\bf 451} (1995) 525.

\bibitem{Green:1995ga}
  M.~B.~Green and M.~Gutperle,
  Nucl.\ Phys.\ B {\bf 460} (1996) 77.

\bibitem{Watson:2004aq}
  S.~Watson,
  Phys.\ Rev.\ D {\bf 70} (2004) 066005.

\bibitem{Kadota:2003tn}
  K.~Kadota and E.~D.~Stewart,
  JHEP {\bf 0312} (2003) 008.

\bibitem{Felder:1999pv}
  L.~Kofman, A.~D.~Linde and A.~A.~Starobinsky,
  Phys.\ Rev.\ D {\bf 56} (1997) 3258;
G.~N.~Felder, L.~Kofman and A.~D.~Linde,
  Phys.\ Rev.\ D {\bf 60} (1999) 103505;
G.~N.~Felder, L.~Kofman and A.~D.~Linde,
  Phys.\ Rev.\ D {\bf 59} (1999) 123523.

\bibitem{Brustein:2002mp}
  R.~Brustein, S.~P.~De Alwis and E.~G.~Novak,
  Phys.\ Rev.\ D {\bf 68} (2003) 023517.

\bibitem{Starobinsky:1994bd}
  A.~A.~Starobinsky and J.~Yokoyama,
  Phys.\ Rev.\  D {\bf 50} (1994) 6357
  [arXiv:astro-ph/9407016].

\bibitem{Vilenkin:1982wt}
  A.~Vilenkin and L.~H.~Ford,
  Phys.\ Rev.\ D {\bf 26} (1982) 1231; A.~D.~Linde,
  Phys.\ Lett.\ B {\bf 116} (1982) 340.

\bibitem{Giddings:2001yu}
  S.~B.~Giddings, S.~Kachru and J.~Polchinski,
  Phys.\ Rev.\ D {\bf 66} (2002) 106006.

\bibitem{Derendinger:1985kk}
  J.~P.~Derendinger, L.~E.~Ibanez and H.~P.~Nilles,
  Phys.\ Lett.\ B {\bf 155} (1985) 65;
E.~Witten,
  Nucl.\ Phys.\ B {\bf 474} (1996) 343.

\bibitem{Dine:1998qr}
  M.~Dine, Y.~Nir and Y.~Shadmi,
  Phys.\ Lett.\ B {\bf 438} (1998) 61;
M.~Dine, L.~Randall and S.~Thomas,
  Phys.\ Rev.\ Lett.\  {\bf 75} (1995) 398.

\bibitem{Greene:2007sa}
  B.~Greene, S.~Judes, J.~Levin, S.~Watson and A.~Weltman,
  arXiv:hep-th/0702220.

\bibitem{Linde:2001ae}
  A.~Linde,
  JHEP {\bf 0111} (2001) 052.

\bibitem{Liddle:2000cg}
  A.~R.~Liddle and D.~H.~Lyth,
 {\em Cosmological Inflation and Large Scale Structure},
(Cambridge University Press, Cambridge U.K., 2000).

\bibitem{BuenoSanchez:2006xk}
  J.~C.~Bueno Sanchez, K.~Dimopoulos and D.~H.~Lyth,
  JCAP {\bf 0701} (2007) 015
  [arXiv:hep-ph/0608299].

\bibitem{Kachru:2003aw}
  S.~Kachru, R.~Kallosh, A.~Linde and S.~P.~Trivedi,
  Phys.\ Rev.\ D {\bf 68} (2003) 046005.

\bibitem{Kaloper:2004yj}
  N.~Kaloper, J.~Rahmfeld and L.~Sorbo,
  Phys.\ Lett.\  B {\bf 606} (2005) 234
  [arXiv:hep-th/0409226].

\bibitem{Hellerman:2001yi}
S.~Hellerman, N.~Kaloper and L.~Susskind,
JHEP {\bf 0106} (2001) 003;
W.~Fischler, A.~Kashani-Poor, R.~McNees and S.~Paban,
JHEP {\bf 0107} (2001) 003;
E.~Witten,
hep-th/0106109.

\bibitem{Copeland:1997et}
E.~J.~Copeland, A.~R.~Liddle and D.~Wands,
Phys.\ Rev.\ D {\bf 57} (1998) 4686.

\bibitem{Cline:2001nq}
  J.~M.~Cline,
  JHEP {\bf 0108} (2001) 035;
C.~F.~Kolda and W.~Lahneman,
hep-ph/0105300.

\bibitem{Blais:2004vt}
  D.~Blais and D.~Polarski,
  Phys.\ Rev.\ D {\bf 70} (2004) 084008.

\bibitem{Giovannini:1999bh}
  M.~Giovannini,
  Phys.\ Rev.\ D {\bf 60} (1999) 123511;
V.~Sahni, M.~Sami and T.~Souradeep,
  Phys.\ Rev.\ D {\bf 65} (2002) 023518.

\bibitem{Biswas:2003kf}
  T.~Biswas and P.~Jaikumar,
  Phys.\ Rev.\ D {\bf 70} (2004) 044011;
T.~Biswas and P.~Jaikumar,
  JHEP {\bf 0408} (2004) 053.

\bibitem{Klebanov:2000hb}
  I.~R.~Klebanov and M.~J.~Strassler,
  JHEP {\bf 0008} (2000) 052.

\bibitem{Burgess:2003ic}
  C.~P.~Burgess, R.~Kallosh and F.~Quevedo,
  JHEP {\bf 0310} (2003) 056.

\bibitem{Becker:2002nn}
  K.~Becker, M.~Becker, M.~Haack and J.~Louis,
  JHEP {\bf 0206} (2002) 060.

\bibitem{Westphal:2005yz}
  A.~Westphal,
  JCAP {\bf 0511} (2005) 003.

\bibitem{Conlon:2005ki}
  J.~P.~Conlon, F.~Quevedo and K.~Suruliz,
  JHEP {\bf 0508} (2005) 007.

\bibitem{LopesFranca:2002ek}
  U.~J.~Lopes Franca and R.~Rosenfeld,
  JHEP {\bf 0210} (2002) 015.

\bibitem{Kallosh:2002gf}
  R.~Kallosh, A.~Linde, S.~Prokushkin and M.~Shmakova,
  Phys.\ Rev.\ D {\bf 66} (2002) 123503.

\bibitem{Kehagias:2004bd}
  A.~Kehagias and G.~Kofinas,
  Class.\ Quant.\ Grav.\  {\bf 21} (2004) 3871.

\bibitem{Turner:1983he}
  M.~S.~Turner,
  Phys.\ Rev.\ D {\bf 28} (1983) 1243.
}
\end{thebiblio}
\end{document}